\documentclass[a4paper,fleqn,usenatbib]{mnras}

\usepackage{newtxtext,newtxmath}

\usepackage[T1]{fontenc}
\usepackage{ae,aecompl}

\usepackage{graphicx}	
\usepackage{amsmath}	
\usepackage{amssymb}	

\title[Charge exchange emission in galactic outflows]
{Charge exchange emission and cold clumps in multi-phase galactic outflows}

\author[K. Wu et al.]{ 
Kinwah Wu$^1$\thanks{E-mail: kinwah.wu@ucl.ac.uk (KW), kaye.li@link.cuhk.edu.hk (KJL), 
  ellis.owen.12@ucl.ac.uk (ERO), ji@pmo.ac.cn (LJ), 
  snzhang@pmo.ac.cn (SZ), 
  g.branduardi-raymont@ucl.ac.uk (GBR)},   
Kaye Jiale Li$^{1,2}$, 
Ellis R. Owen$^{1}$,  
Li Ji$^{3,4}$,
Shuinai Zhang$^{3,4}$  
\newauthor
 \;\! and Graziella Branduardi-Raymont$^{1}$  \vspace*{0.25cm} \\ 
$^{1}$Mullard Space Science Laboratory, University College London, Holmbury St Mary, Surrey RH5 6NT, UK\\
$^{2}$Department of Physics, Chinese University of Hong Kong, Shatin, NT, Hong Kong SAR, China  \\ 
$^{3}$Purple Mountain Observatory, Chinese Academy of Sciences, 8 Yuanhua Road, Nanjing 210034, China \\ 
$^{4}$Key Laboratory of Dark Matter and Space Astronomy, 
  Chinese Academy of Sciences, Nanjing 210034, China
}

\date{Accepted 2019 November 21. Received 2019 November 20; in original form 2018 December 23}

\pubyear{2015}

\begin{document}
\label{firstpage}
\pagerange{\pageref{firstpage}--\pageref{lastpage}}
\maketitle

\begin{abstract}  
Large-scale outflows 
  from starburst galaxies are multi-phase, multi-component fluids. 
Charge-exchange lines which originate from the interfacing surface 
  between the neutral and ionised components  
  are a useful diagnostic of the cold dense structures 
  in the galactic outflow. 
From the charge-exchange lines observed in the nearby starburst galaxy M82, 
  we conduct surface-to-volume analyses 
  and deduce that the cold dense clumps in its galactic outflow 
  have flattened shapes, resembling a hamburger or a pancake morphology 
  rather than elongated shapes. 
The observed filamentary H$\alpha$ features are therefore 
  not prime charge-exchange line emitters. 
They are stripped material torn 
  from the slow moving dense clumps 
  by the faster moving ionised fluid 
  which are subsequently warmed and stretched into elongated shapes.  
Our findings are consistent with numerical simulations 
  which have shown that 
  cold dense clumps in galactic outflows 
  can be compressed by ram pressure, 
  and also progressively ablated and stripped before complete disintegration. 
We have shown that some clumps could survive their passage along a galactic outflow. 
  These are advected into the circumgalactic environment, where their remnants would seed condensation of the circumgalactic medium
  to form new clumps.  
The infall of these new clumps back into the galaxy 
  and their subsequent re-entrainment into the galactic outflow 
  form a loop process of galactic material recycling. 
 
\end{abstract}    
\begin{keywords}
galaxies:starburst -- galaxies:individual(M82) -- X-rays:galaxies -- ISM:structure -- atomic processes 
\end{keywords}



\section{Introduction}

The Universe is mostly filled with ionised gas. 
However, in many astrophysical systems 
  such as galaxies, stars, and sub-stellar objects, 
  neutral materials are not only present but coexist with hot ionised gases. 
In the interfaces where neutral and ionised media meet, 
  ions would inevitably encounter neutral atoms or molecules     
  and the interactions between them would allow for the capture of an electron or several electrons by the ions. 
In this charge-exchange (CX) process,  
  the electron-capture hosts are often in an excited state.   
Their subsequent de-excitation gives rise to photon emission,  
  with the photon energy determined by the energy levels involved in the electronic transition.  
X-ray emission lines are produced when the emitting hosts are highly charged ions.  
CX emission processes are efficient,    
   as the interaction cross-section may reach $10^{-15}{\rm cm}^2$ 
    \citep{Tawara1985,McGrath1989,Greenwood2000}.  
  This is much larger than the Thomson electron-photon scattering cross-section  
  ($\sigma_{\rm T} = 6.65\times 10^{-25}{\rm cm}^2$),   
  which signifies common radiative processes that occur in hot astrophysical plasmas 
  (e.g., accretion flows in compact objects and relativistic AGN jets)  
  which generate keV X-rays.   
CX emission has been observed in astronomical environments ranging from sub-stellar sized objects,  
  e.g.\ comets \citep{Lisse1996,Cravens1997,Bodewits2007},  
  planets \citep{Branduardi-Raymont2004,Dennerl2006,Dennerl2008} 
  and moons \citep{Johnson1982,McGrath1989} 
  to systems of galactic-scale structures and beyond, 
  e.g.\ high-velocity clouds in galactic halos \citep{Lallement2004}, 
  the galactic centre and the galactic ridge  \citep{Tanaka1999,Tanaka2002}, 
  and large-scale outflows from star-forming galaxies  
  \citep{Ranalli2008,Konami2011,Liu2011,Zhang2014}. 
Although CX processes are often associated with ionised or partially ionised gases, 
  a thermally hot medium is not always required.
  An example of this is the solar wind and comet interface \citep[see][]{Ip1989}. 

The current spectroscopic diagnoses of keV and sub-keV astrophysical plasmas 
  are mostly based on the radiative recombination emission 
  associated with collisional ionisation/excitation and photo-ionisation processes 
  \citep[see][]{Smith2001,Ferland2003,Raymond2005,Kaastra2008,Porquet2010}.    
Despite this, CX spectroscopy has been applied in many other disciplinary areas for decades, 
  e.g.\ in confined fusion reaction and reactor studies 
  \citep[e.g.][]{Isler1994,Ida2008,Li2016}.   
  Its use as a diagnostic tool of astrophysical plasmas outside the solar system~\citep[see][]{Wargelin2008,Dennerl2010} 
  has also recently gained attention.   
CX lines are now identified as a useful means 
  by which the physical conditions in multi-phase outflows from galaxies can be probed 
  \citep[see][]{Konami2011,Liu2011,Liu2012MNRAS,Zhang2014}.   

Large-scale galactic outflows are characteristic of starburst galaxies 
  \citep[see][for reviews]{Heckman2003,Veilleux2005}. 
They are powered by the star-forming processes in galaxies  
  \citep{Mathews1971,Chevalier1985,Strickland2000,Meiksin2016}, 
  where the frequent supernova explosions continually inject 
  large amounts of energy into the interstellar medium (ISM).  

The out-of-galactic-plane outflow material emits strong continuum and line X-rays 
  \citep{Bregman1995,Dahlem1998,Wang2001,Strickland2004,Yamasaki2009},
  implying that it contains substantial hot, thermal, ionised gases.  
The discovery of synchrotron emission halos associated with galactic outflows  
  in some starburst galaxies, e.g.\ NGC~253 \citep{Carilli1992}, 
 also indicate the presence of energetic non-thermal electrons and cosmic rays.  
However, the filament-like structures in H$\alpha$ images \citep[see][]{Lehnert1996,Hoopes1999,Yoshida2011},  
  suggest that warm, partially ionised gas is interwoven amongst the hot ionised keV X-ray emitting gases.  
Colour variations in the optical/UV emission are observed 
  in the conic outflows (hereafter, wind cone) 
  of galaxies \citep{Hutton2014,Hutton2015}, 
  which are attributed to differential extinction caused by variations 
  in the dust properties and/or temperature inhomogeneity along the flows. 
There is an ionisation cap at high altitudes ($\sim 10~{\rm kpc}$) 
   of the outflow in some starburst galaxies, e.g., M82  
   \citep{Devine1999,Tsuru2007}.     

Although cooling effects are arguably very significant in altering the flow dynamics 
  in the so-called super-winds of starburst galaxies \citep{Heckman2003}, 
  the thermal properties in the outflow 
  are by no means uniform throughout   
  (see e.g., the thermal and hydrodynamic profiles shown in \citealt{Chevalier1985}). 
Recombination and radiative cooling would compete with ionisation and mechanical (shock) heating 
  in the wind cone \citep{Hoopes2003}, the transitional cap\footnote{This 
    transitional cap is not the ionisation cap mentioned above, 
    but a rough boundary 
    where there is a transition between the properties 
    of the inhomogeneities in the galactic wind cone.}, and the region above it. 
  Moreover, the thermal and dynamical instabilities that subsequently develop  
  will lead to the fragmentation of the ionised flow into ionised bubbles interspersed with cooler, denser, condensed clumps 
  \citep[see e.g.][]{Strickland2000,Pittard2003,Cooper2008ApJ,Fujita2009ApJ}.  
These ionised and neutral sub-structures are entrained 
   in the flow
   and continue migrating upwards \cite[see][]{Schwartz2004}.   
As for the hot ionised bubbles, 
  some will escape to intergalactic space 
  where gravitational forces give way to inertial, radiative and buoyancy forces 
  \citep[cf. the model of bubble buoyancy in Cen A,][]{Saxton2001}, while others may be heated and evaporated as they traverse up the outflow zone.   
Galactic outflows are evidently complex, multi-component, multi-phase fluids 
  \citep[see][]{Ohyama2002,Strickland2002,Melioli2013,Martin-Fernandez2016},   
  where hot ionised gases, warm partially ionised gases and cool neutral material 
  intermingle as well as segregate.   

While we are able to paint a broad phenomenological picture of galactic  outflows, 
  many questions regarding their finer geometries and physical properties 
  are still waiting to be answered. 
For instance, what is the mass-loading of neutral material and warm gas 
  in the hot ionised outflows? 
What are the filling factors of the ionised and neutral components in flows? 
What are the internal geometries of the neutral material and the warm gas?  
The answers to these questions are not only essential 
  to determine the dynamics of the system,  
  but they also provide insights for wider astrophysical issues
  such as the chemical and energy transport processes at work within galaxies, and from galaxies into intergalactic space, 
  and the role of galactic outflows in shaping the present-day structure of our Universe. 

In this work we investigate the interior geometries of 
  multi-phase multi-component galactic outflows, 
  utilising the information obtained from the CX spectroscopic analyses.  
We determine the surface area to volume ratio of ionised gas and neutral material in a flow,  
  which characterises the strength of the CX emission, 
  and we derive the volume partitions of neutral and ionised fluid components.  
We show that CX emission information 
  together with appropriate models for the thermal and dynamical properties 
  of the ionised gas and neutral material  
  can strongly constrain the internal geometries of outflow regions in starburst galaxies. 

We organise the paper as follows: 
In \S 2 we present a two-phase two-zone model 
  for the outflow region from star-forming galaxies, 
  with spherical neutral clumps entrenched in an ionised zone 
    and spherical gas bubbles in a neutral zone, 
    and we compute the surface-to-volume ratio of the ionised gas.   
In \S 3 we relax the model to allow ellipticity in the neutral clumps and the ionised gas bubbles 
   and, finally, we generalise the model to consider clumps and bubbles with arbitrary aspect ratios. 
We compute the corresponding surface-to-volume ratios for these cases. 
With these ratios in-hand,  
  we use values inferred from the charge-exchange (CX) lines 
  observed in galactic outflows of starburst galaxies 
  and the timescale over which condensed, neutral clumps are ablated  
  to set constraints on the geometries and sizes of neutral clumps
  and their filling factors within ionised outflows.  
In \S4 we discuss our findings 
  in comparison to previous numerical and observational studies. 
We also discuss the astrophysical implications 
  in the context of the survival of the neutral clumps in the galactic outflows 
  and the advection of remnant neutral clumps and their stripped gas 
  into the circumgalactic medium.    
In \S 5 we give a brief summary of our findings. 

\section{Two-phase two-zone model} 

Consider that the outflow region is conical, and is enclosed in a neutral background medium.   
The outflow region consists of two zones, 
 the first being predominantly ionised, and the other being predominantly neutral. 
The predominantly ionised zone (hereafter the ionised zone) is located close to the galactic plane, 
  and the predominantly neutral zone (hereafter the neutral zone) is above the ionised zone at the outskirts.  
Both zones are inhomogeneous. 
In the ionised zone, 
 neutral clumps are entrained in an ionised flow; 
in the neutral zone, ionised bubbles 
 are embedded in the neutral outflowing material. 

Without losing generality we assume that the conic outflow region has a circular cross-section.  
The opening half-angle of the wind cone is $\alpha$, and the height is $h$. 
The ionised zone terminates at a height $h' (=\chi h\, , \,  \chi \leq1)$, 
   and above it lies the neutral zone. 
The radius of the circular surface where the two zones meet 
  is $h' \tan\alpha$. 

\subsection{Surface area to volume ratios of individual spherical clumps and bubbles} 

\begin{figure}
\begin{center} 
\hspace*{-0.7cm}
\includegraphics[width=10cm]{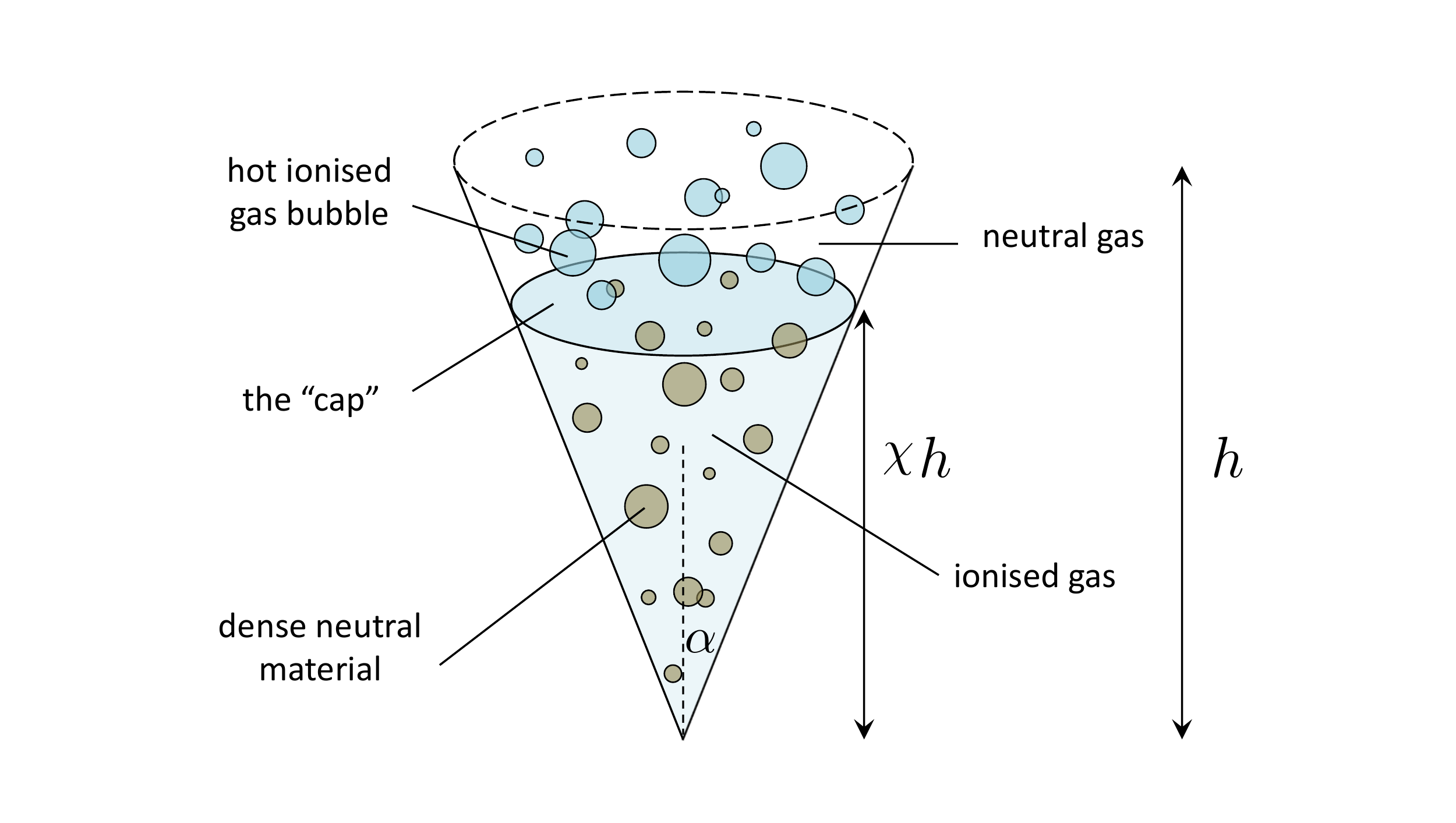}
\end{center} 
\vspace*{-0.2cm}
\caption{A schematic illustration of the two-component, two-phase wind cone model 
       used in this study. 
    The ionised gas occupies the bottom region, 
    and neutral clumps are embedded within the ionised gas. 
    The region above the ionised gas cone is occupied by predominantly neutral material. 
    Bubbles of hot ionised gas are present in this neutral region. 
    The ``cap" is the transitional layer 
    between the ionised 
    region at the bottom of the outflow cone
      and the neutral material at the top.  
    The entire wind cone is enclosed by neutral material (not shown). 
    Although the clumps and bubbles are drawn as spheres in the diagram, 
      they may assume any shape and various aspect ratios 
      within the appropriate astrophysical contexts of the calculations. 
    The opening half-angle of the wind cone is $\alpha$,  
    and the height of the entire cone is $h$. 
       The fractional height of the bottom ionised region is specified by the parameter $\chi$.  
       }
    \label{fig:wind_cone}
\end{figure}

We first consider that both the dense neutral clumps in the ionised zone 
  and the ionised bubbles in the neutral zone are spherical. 
The spherical assumption will be relaxed in later sections, 
  where other shapes will be considered.     
In reality, 
  the clumps and the bubbles 
  would vary in their sizes and densities. 
Without losing generality, 
  we assume that 
  the sizes and the densities 
  of the clumps and bubbles 
  can be represented by some characteristic values. 
Then we may assign an effective radius $r_1$ 
  for the neutral clumps 
  and an effective radius $r_2$ for the ionised bubbles, 
  and similarly 
  an effective number density $\rho_1$ for the clumps  
  and an effective number density $\rho_2$ for the bubbles.  
Moreover,   
  the sizes of the neutral clumps and the ionised gas bubbles 
  are small, with the condition 
  ${\rm Max}\;\!\!(r_1, r_2) \ll {\rm Min}\;\!(\chi h\tan\alpha,\chi h,h(1-\chi))$ 
  generally satisfied. 
A schematic illustration of the two-zone wind cone model 
  is shown in Fig.~\ref{fig:wind_cone}. 

The total volume of the outflow region is $V_{\rm w}  =  (\pi h^3 \tan^2\! \alpha)/3$, 
  while the volumes of the ionised and neutral zones are $\chi^3 V_{\rm w}$ 
    and $(1- \chi^3) V_{\rm w}$ respectively. 
The total volume occupied by the ionised gas in the entire outflow is 
\begin{equation}  
  V_{\rm ion} = V_{\rm w} 
    \left\{\chi^3\left[1 - \rho_1 \bigg(\frac{4\pi}{3} r_1^3 \bigg)\right]   
       +(1-\chi^3) \rho_2 \bigg(\frac{4\pi}{3} r_2^3 \bigg) \right\}   \ , 
\end{equation} 
  and the total volume occupied by the neutral material is 
\begin{equation} 
  V_{\rm neu} =  V_{\rm w} 
    \left\{(1-\chi^3)\left[1 - \rho_2 \bigg(\frac{4\pi}{3} r_2^3 \bigg)\right]   
       +\chi^3 \rho_1 \bigg(\frac{4\pi}{3} r_1^3 \bigg) \right\}  \ . 
\end{equation}  

There are four interfaces between the ionised gases and the neutral material in this configuration: 
(i) the interface at which the ionised zone and the neutral zone 
  within the outflow region meet; 
(ii) the interface at which the ionised zone meets with the neutral material enclosing the wind cone; 
(iii) the interface between the ionised gas and the dense neutral clumps in the ionised zone; and 
(iv) the interface between the neutral material and the ionised gas bubbles in the neutral zone. 
If the interfacing surfaces in all these cases are smooth, 
  the respective areas are simply  
  $\pi r^2\, (= \chi^2 \pi h^2 \tan^2\!\alpha)$, 
  $\pi r \sqrt{r^2 +h'^2}\,(= \chi^2 \pi h^2 \tan^2\!\alpha\, {\rm cosec}\,\alpha )$, 
  $\chi^3 V_{\rm w} (\rho_1 4\pi r_1^2)$ and $(1- \chi^3) V_{\rm w} (\rho_2 4\pi r_2^2)$. 
Summing them gives the total surface area of the interfacing boundaries 
   between the ionised gas and the neutral material in the outflow:  
\begin{multline} 
  A_{\rm ion} =   A_{\rm w} \bigg\{  \chi^2 
     + \frac{h}{r_1} \frac{\chi^3}{(1+ {\rm cosec}\, \alpha)}  \rho_1 \bigg( \frac{4\pi}{3} r_1^3\bigg) \\
      + \frac{h}{r_2} \frac{1- \chi^3}{(1+ {\rm cosec}\, \alpha)} \rho_2 \bigg( \frac{4\pi}{3} r_2^3\bigg)
    \bigg\}  \ , 
\end{multline} 
  where the total surface area of the outflow zone is 
\begin{equation} 
   A_{\rm w} = {\pi}\, h^2 \tan^2\alpha (1+ {\rm cosec}\, \alpha) \ . 
\end{equation} 

Note that $\rho_1 (4\pi r_1^3)/3 \; (=f_1)$ and $\rho_2 (4\pi r_2^3)/3 \; (=f_2)$ 
   are the volume filling fraction of the dense neutral clumps in the ionised zone 
   and the volume filling fraction of the ionised gas bubbles in the neutral zone respectively.  
The relative enhancement of the area-to-volume ratio 
  for the surface interfacing the ionised gases and the neutral material 
  in the presence of spherical dense neutral clumps in the ionised zone 
  and spherical ionised gas bubbles in the neutral zone is therefore 
\begin{equation} 
\begin{split} 
{\cal R}  & =  \frac{A_{\rm ion}/V_{\rm ion}}{A_{\rm w}/V_{\rm w}} \\
 &  = \frac{\chi^2(1+{\rm cosec} \alpha) + \frac{h}{r_1} {\chi^3 f_1} 
        +  \frac{h}{r_2} {(1-\chi^3) f_2}}{
         (1+{\rm cosec} \alpha)\left[\chi^3(1-f_1)+(1-\chi^3)f_2\right]} \ . 
\end{split} 
\end{equation} 

In the special case where the outflow is predominantly occupying
  the ionised zone (with clumps of neutral material), $\chi =1$ and 
\begin{equation} 
  {\cal R} = \frac{(1+{\rm cosec}\,\alpha)+\frac{h}{r_1} f_1}{(1+{\rm cosec}\,\alpha)(1-f_1)} \ . 
\end{equation} 
If the volume filling fraction of the dense neutral clumps is insignificant, 
  i.e.\ $h f_1 \ll r_1$  (which automatically implies $f_1 \ll 1$, although the converse is not always true),  
  the global geometrical factor $(1+ {\rm cosec}\, \alpha)$ will dominate in the numerator. 
Then, we have 
\begin{equation} 
  {\cal R} \approx 1 \ .   
\end{equation}  
This recovers the result for a fully ionised outflow enveloped by a neutral ambient medium.   
If the volume filling fraction of the dense neutral clumps is not negligible, 
  and if the clump sizes are sufficiently small such that $r_1 \ll h$, then 
\begin{equation} 
  {\cal R}  \approx \frac{h}{r_1}\left[ \frac{f_1}{(1+{\rm cosec}\,\alpha)(1-f_1)} \right] \ . 
  \label{eq:r_factor_obs1}
\end{equation}    

\subsection{Surface area to volume ratios of individual ellipsoidal clumps and bubbles}

We now relax the spherical assumption for the dense neutral clumps and ionised gas bubbles.    
Consider that they are ellipsoids, characterised by three semi-axes $a$, $b$ and $c$. 
The volume of these ellipsoids is simply   
\begin{equation} 
  V_{\rm ep} = \frac{4\pi}{3} abc  \ . 
\end{equation}  
One needs to evaluate two elliptic integrals numerically to obtain the exact surface area of an ellipsoid.     
A simple analytical expression in terms of elementary functions is not always possible.  
This is impractical when extracting information from observational data 
   where an algebraic expression of the surface area is unavailable. 
An approximate analytic expression with good accuracy is therefore needed. 
Hence, we may adopt the Thomsen formula (see Appendix \ref{A:ep})
  for the surface area of ellipsoids   
\begin{equation} 
 A_{\rm ep} = 4\pi \left[ \frac{1}{3} (a^p b^p + b^p c^p + c^p a^p)\right]^{1/p} \ ,  
\end{equation}   
  with the index $p \approx 1.6075$. 
Then we obtain a surface area to volume ratio for the ellipsoids as:   
\begin{equation} 
  \frac{A_{\rm ep}}{V_{\rm ep}} = 
    3 \left[\frac{1}{3}\bigg(\frac{1}{a^p} +\frac{1}{b^p} +\frac{1}{c^p}   \bigg) \right]^{1/p}  \ . 
\end{equation} 
The surface area to volume ratio of a sphere with a radius $r$ is 
\begin{equation} 
  \frac{A_{\rm sp}}{V_{\rm sp}} = \frac{3}{r}  \ .
\end{equation} 
Thus, the equivalent radius of a sphere with a volume the same as that of an ellipsoid 
   is the geometric mean of the three semi-axes of the ellipsoid, i.e., 
\begin{equation} 
  r = (abc)^{1/3} \ . 
\end{equation}
       
We define the ratio $\Upsilon = A_{\rm ep}/A_{\rm sp}$ of the surface areas of the ellipsoid and the sphere respectively. 
For an ellipsoid and a sphere with the same volume, 
\begin{equation} 
\begin{split} 
  {\hat \Upsilon} & = \frac{A_{\rm ep}/V_{\rm ep}}{A_{\rm sp}/V_{\rm sp}} 
     \bigg\vert_{V_{\rm ep}=V_{\rm sp}}\\ 
     & = r \left[\frac{1}{3}\bigg(\frac{1}{a^p} +\frac{1}{b^p} +\frac{1}{c^p}   \bigg) \right]^{1/p} \\ 
     & = \left[\frac{1}{3}\bigg(\frac{1}{a^p} +\frac{1}{b^p} +\frac{1}{c^p}   \bigg)
      \bigg(\frac{1}{a^p}\frac{1}{b^p}\frac{1}{c^p} \bigg)^{-1/3} \right]^{1/p} \ .   
\end{split} 
\label{eq:ellipsoids}
\end{equation} 
The $p$-th power of ${\hat \Upsilon}$ 
   is essentially the ratio between the arithmetic mean and the geometrical mean 
   of the reciprocals of the three semi-axes. 
It is always equal to or larger than 1, and is equal to one only when $a=b=c$ (i.e. a sphere).  
${\hat \Upsilon}$ serves as a geometrical correction factor 
  for the surface area of an object when it deviates from a spherical shape. 
Thus, a general formula for the enhancement of the surface area to volume ratio 
  of the ionised material, taking into account the ellipsoid shapes of the dense neutral clumps  
  and the ionised gas bubbles, is 
\begin{equation} 
{\cal R}  =  \frac{\chi^2(1+{\rm cosec} \alpha) + \frac{h}{r_1} {\chi^3 {\hat \Upsilon}_1 f_1} 
        +  \frac{h}{r_2} {(1-\chi^3) {\hat \Upsilon}_2 f_2}}{
         (1+{\rm cosec} \alpha)\left[\chi^3(1-f_1)+(1-\chi^3)f_2\right]} \ ,  
\label{eq:RE}
\end{equation} 
 where $\{{\hat \Upsilon_1(a_1,b_1, c_1)}, {\hat \Upsilon}_2(a_2, b_2, c_2)\}$ 
  are the geometrical correction factors
   for the ellipsoidal dense neutral clumps and the ionised gas bubbles, 
   and here $r_1 = (a_1 b_1 c_1)^{1/3}$ and  $r_2 = (a_2 b_2 c_2 )^{1/3}$.  

\subsection{Effects of the aspect ratios of individual clumps and bubbles}  

\subsubsection{Cylindrical approximation} 

Now consider that the clumps and bubbles 
  have shapes with substantial aspect ratios. 
Here we adopt a simple approximation that 
  the elongated and flattened clumps (hereafter filament-like and pancake-like correspondingly) 
  and bubbles are approximated by cylinders.   
The height/length of the cylinder is $t$ and the cross-sectional area is $\pi s^2$ 
  (where $s$ is the radius of the cross-section).  
For the filament-like clumps/bubbles, $t \gg s$;   
 for the pancake-like clumps/bubbles $t \ll s$.  
The surface area of $2\pi s(s+t)$ and the volume of $\pi s^2 t$ of the cylinders 
  a geometrical correction factor 
\begin{equation} 
\begin{split} 
  {\hat \Upsilon} & = \frac{A_{\rm ep}/V_{\rm ep}}{A_{\rm sp}/V_{\rm sp}} 
     \bigg\vert_{V_{\rm ep}=V_{\rm sp}}\\ 
     & =  \frac{2}{3} \left[ \frac{r \;\!(s+t)}{s\;\!t}\right] \\ 
     & =  \left(\frac{2}{9} \right)^{1/3} \left[\frac{(s+t)}{s^{1/3} t^{2/3}} \right]   \ .   
\end{split} 
\end{equation} 
Now consider an aspect ratio defined as   
\begin{equation} 
  \zeta \equiv   \frac{{\rm Max}\,(s,t/2)}{{\rm Min}\,(s,t/2)} \ . 
\end{equation} 
In terms of this aspect ratio, the geometrical correction factor is
\begin{equation} 
  {\hat \Upsilon} 
   =  \left(\frac{1}{18} \right)^{1/3} \left[   \frac{n+(3-n)\zeta}{\zeta^{(3-n)/3}}  \right] \ , 
\label{eq:n1n2}
\end{equation} 
   with $n=1$ for filament-like clumps and bubbles 
   and $n=2$ for pancake-like clumps and bubbles. 
In the limit of an extreme aspect ratio, the expression becomes   
\begin{equation} 
  {\hat \Upsilon} 
   \approx  \left(\frac{1}{18} \right)^{1/3}\!\!(3-n)\, \zeta^{n/3}   \ .  
\end{equation}  

In the cylindrical approximation, 
  the enhancement of the surface area to volume ratio ${\cal R}$
  has the same expression as that in equation \ref{eq:RE}, 
  except that the equivalent spherical radius is now 
\begin{equation}  
  r = \left(\frac{3}{2}\right)^{1/3} \frac{t}{2} \  \zeta^{-2/3}\   
\end{equation}  
  for the filament-like clumps and bubbles and 
\begin{equation}  
  r = \left(\frac{3}{2}\right)^{1/3} s \  \zeta^{-1/3}\   
\end{equation}  
  for the pancake-like clumps and bubbles. 
It is interesting that, in the limit of extreme aspect ratios for the clumps or bubbles, 
  i.e. ${\zeta^{-1}\rightarrow 0}$, the factor is therefore 
\begin{equation}  
   \frac{h}{r} {\hat \Upsilon} \approx  \frac{2}{3} \left[  \frac{2 h\;\!  \zeta }{t} \right]
     =  \frac{2}{3}  \frac{h}{s}  \ ,
\end{equation}   
   for filament-like cylindrical clumps and bubbles and 
\begin{equation}  
   \frac{h}{r} {\hat \Upsilon} \approx  \frac{2}{3} \left[  \frac{h\;\! \zeta }{2s} \right]
     =  \frac{1}{3}  \frac{h}{t/2}  \ ,
\end{equation}   
  for pancake-like cylindrical clumps and bubbles.  
   
\subsubsection{Elongated and flattened ellipsoids} 

Here we consider the case where the clumps/bubbles are elongated and flattened ellipsoids.  
For ellipsoids with a rotational symmetry axis, 
  the geometrical correction factor ${\hat \Upsilon}$, in equation \ref{eq:ellipsoids} obtained previously, 
  can be expressed in terms of an aspect ratio, which is now defined as  
\begin{equation} 
 \zeta \equiv \frac{{\rm Max}\,(a,b,c)}{{\rm Min}\,(a,b,c)} \ . 
\end{equation} 
In the convention $a \geq b \geq c$, $\zeta$ is simply $a/c$.   
Setting $a> b = c$ gives       
\begin{equation} 
 {\hat \Upsilon} = \left[ \frac{1}{3} \bigg( \zeta^{-2p/3} + 2 \zeta^{p/3} \bigg) \right]^{1/p}      
\end{equation} 
 for the prolate ellipsoids, and setting $a = b > c$ yields  
\begin{equation} 
 {\hat \Upsilon} =  \left[ \frac{1}{3} \bigg(2 \zeta^{-p/3} +  \zeta^{2p/3} \bigg) \right]^{1/p}     
\end{equation}  
  for the oblate ellipsoids.  
Filaments can be considered as prolate ellipsoids with $\zeta \gg 1$, 
  and their geometrical correction factor is therefore  
\begin{equation} 
   {\hat \Upsilon} \approx \left( \frac{2}{3}\right)^{1/p} \zeta^{1/3}   \ . 
\end{equation}  
Oblate ellipsoids with $\zeta \gg 1$ would resemble pancakes, 
  and their geometrical correction factor is 
\begin{equation} 
   {\hat \Upsilon} \approx \left( \frac{1}{3} \right)^{1/p}   \zeta^{2/3}   \ . 
\end{equation}  
The effective spherical radii $r$ of the ellipsoids are  
\begin{equation} 
  r (n) = a \zeta^{-(3-n)/3}  \ ,
\end{equation} 
   with $n =1$ for prolate ellipsoids and $n=2$ for oblate ellipsoids.  
Hence, 
\begin{equation} 
    \frac{h}{r} {\hat \Upsilon} \approx 
      \left(\frac{3-n}{3}\right)^{1/p}\! \frac{h\, \zeta}{a} =   \left(\frac{3-n}{3}\right)^{1/p}\! \frac{h }{c} \ ,   
\end{equation} 
  for the ellipsoids. 
It is also worth noting that $(1/3)^{1/p} \approx 0.5049$ and $(2/3)^{1/p} \approx 0.7771$ 
  for $p = 1.6075$ (see Appendix \ref{A:ep}), 
  which implies that $1/3<(1/3)^{1/p} < 2/3 < (2/3)^{1/p}$.  

\begin{figure}
\begin{center} 
\hspace*{-0.0cm}
\includegraphics[width=8.2cm]{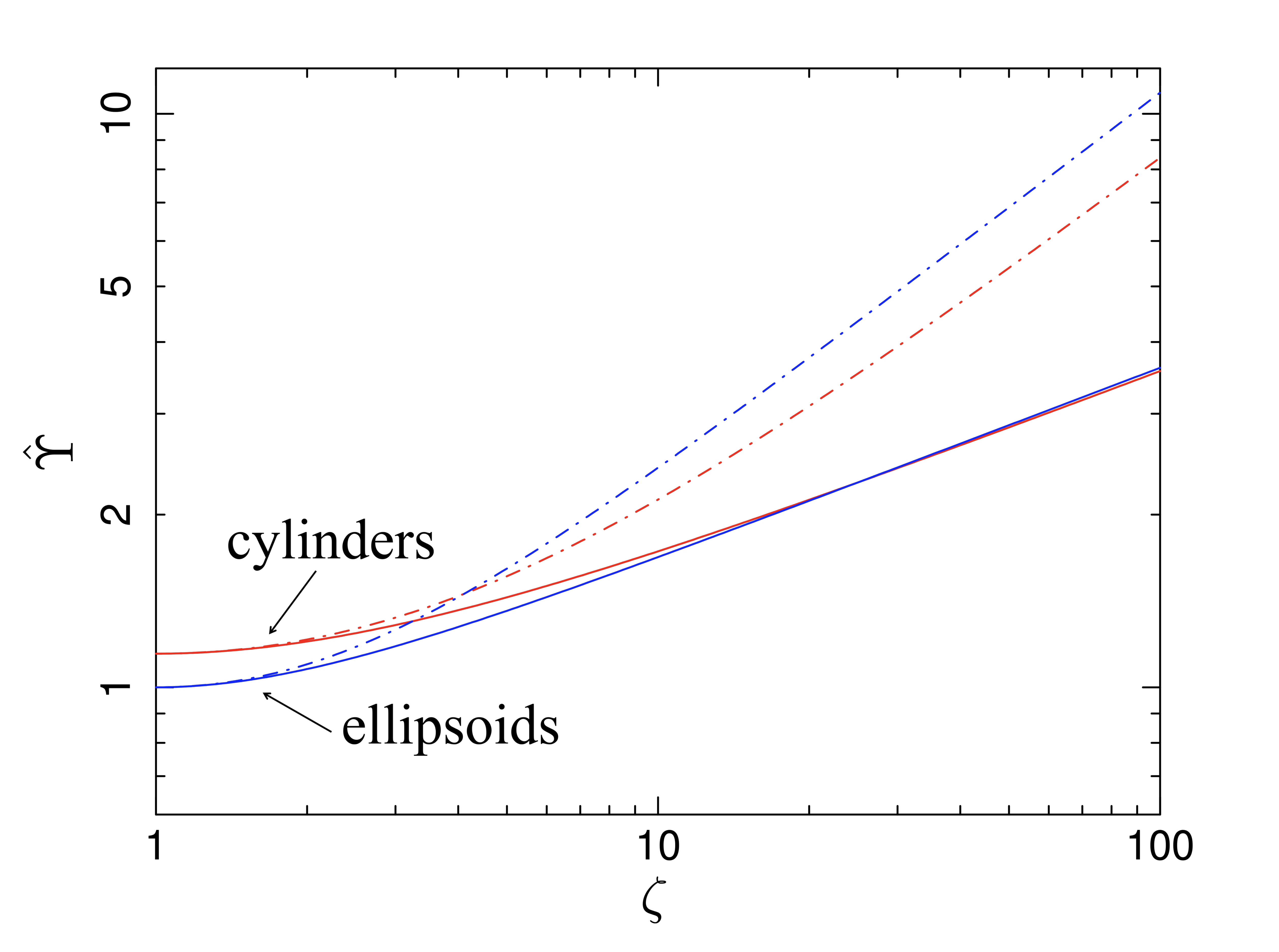}
\end{center} 
\vspace*{-0.5cm}
\caption{Plot of the surface area enhancement factor ${\hat \Upsilon}$ 
     against the aspect ratios $\zeta$ of cylindrical and axisymmetric ellipsoidal clumps/bubbles.   
  The solid lines denote long cylinders (corresponding to $n=1$ in equation \ref{eq:n1n2}) 
    and prolate ellipsoids (corresponding to the filament-like clumps/bubbles), while  
    the dot-dashed lines denote short cylinders (corresponding to $n=2$ in equation \ref{eq:n1n2}) 
    and oblate ellipsoids (corresponding to the pancake-like clumps/bubbles). }
\label{fig:head}
\end{figure}

Fig.~\ref{fig:head} shows the dependence of ${\hat \Upsilon}$ on $\zeta$ 
   for elongated and flattened clumps/bubbles modelled respectively by cylinders and ellipsoids.  
As expected, the ellipsoids give $\hat \Upsilon =1$   
    and the cylinders give $\hat \Upsilon =(3/2)^{1/3}$ at $\zeta = 1$.  
In both the elongated and flattened cases,     
  ${\hat \Upsilon}$ of the ellipsoids and the cylinders have the same dependencies on $\zeta$ 
  when the aspect ratio is sufficiently large. 
Moreover, the value of $\hat \Upsilon$ hardly exceeds 12, even when $\zeta$ is as large as 100.  
This restricts the values for the ratios of the interfacing surface areas 
   to the volumes between the ionised and neutral material. 
Nevertheless, higher surface area to volume ratios can be obtained  
  if the interacting boundary is sufficiently rough (e.g.\ as would be the case with fractal structures). 
  
  Shapes with small aspect ratios may also emerge in outflows: for instance, clumps with a substantial relative speed with respect to the velocity 
  of the outflowing plasma 
  in which they are entrained could develop into shapes resembling a hamburger.  We discuss the resulting surface area to volume ratios of such objects in Appendix~\ref{A-one}.

\section{Results and discussion}  

The general two-zone flow model is characterised by $\chi$, which takes a value between 0 and 1, 
with the general formula for the enhancement of the surface area to volume ratio ${\cal R}$ being described by equation~\ref{eq:RE}.
We demonstrate our model by considering the two extreme cases in which the two-zone model reduces to a one-zone system:
(1) when $\chi =0$, in which a neutral background flow is embedded with ionised bubbles;
and (2) when $\chi =1$, in which an ionised flow is embedded with neutral clumps.
A more thorough parameter study and investigation of the two-zone model (i.e. when adopting values of $\chi$ between 0 and 1), is left to future work.

\subsection{Ionised bubbles in a neutral flow} 

With the surface area to volume ratio of individual neutral clumps ${\hat \Upsilon}_1$ 
  and ionised bubbles ${\hat \Upsilon}_2$
  for specific shapes determined, 
  we are ready to calculate the enhancement of the surface area to volume ratio ${\cal R}$.   
The relative size of the ionised gas dominated region 
  and the neutral material dominated region in the wind cone is specified by $\chi$. 

\begin{figure}
\begin{center} 
\hspace*{-0.0cm}
\includegraphics[width=8.2cm]{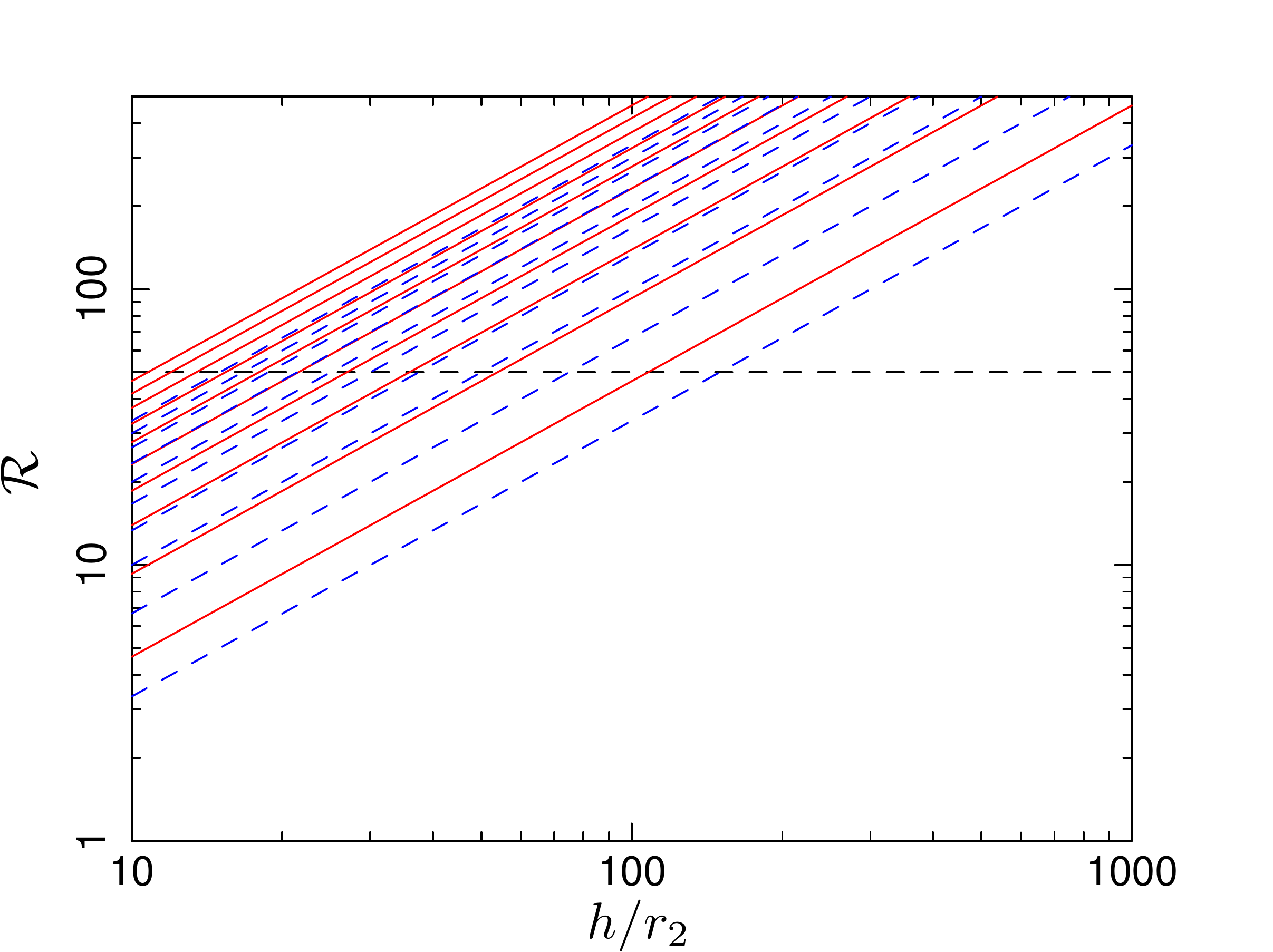} 
\end{center} 
\vspace*{-0.0cm}
\caption{The relative enhancement of the surface area to volume ratio ${\cal R}$ 
   of a neutral outflow with embedded ionised gas bubbles    
     as a function of bubble size, in terms of $h/r_2$, 
     with respect to the height $h$ of the wind cone (calculated from equation~\ref{eq:RE-c0}).  
   The opening half-angle of the conic wind zone  $\alpha = 60^\circ$ (solid curves) and $30^\circ$ (dashed curves),  
     and the parameter $\chi =0$.   
  The curves from top to bottom in each panel 
     correspond to ${\hat \Upsilon} = 10,$ 9, 8, 7, 6, 5, 4, 3, 2, and 1 respectively.     
  The region below the lowermost curve (with ${\hat \Upsilon} = 1$) in each case is forbidden  
     as no physical object has a surface area to volume ratio smaller than that of a sphere.   
A horizontal (dashed) line with ${\cal R}=50$ is shown as a reference. 
This value of ${\cal R}$ is similar to that derived from the CX line spectroscopy (${\cal R}= 27 $) of the superwind of the starburst galaxy M82 -- see section~\ref{sec:calc_r_obs_m82} and~\citet{Zhang2014}.}
\label{fig:ee00}
\end{figure}

We first consider the case with $\chi =0$ 
  in which the wind cone is one-zone, with a neutral background flow which is embedded with ionised bubbles. 
Putting $\chi = 0$ in equation (\ref{eq:RE}) gives   
\begin{equation} 
  {\cal R}  =   \left(\frac{h}{r_2} \right) \frac{{\hat \Upsilon}_2}{(1+{\rm cosec} \alpha)}  \ .   
\label{eq:RE-c0}
\end{equation} 
The enhancement factor ${\cal R}$ thus depends only on 
  the size ($r_2$) and the shape (through ${\hat \Upsilon}$) of the bubbles, 
  when the opening angle of the wind cone is specified.     
Hence, it can be used 
   to put constraints on the geometrical properties of the bubbles in a neutral outflow. 
For instance, in a predominantly neutral flow (i.e.\ with $\chi \approx 0$),  
  a value of ${\cal R} \approx 50$
 derived from CX line spectroscopy in an observation 
  would imply (from Fig.~\ref{fig:ee00}) that the effective linear size of the ionised bubbles, $r_2$,  
  relative to the height of the wind cone, $h$, is about 1/200,   
 provided that the bubbles are sufficiently hot so as to maintain a roughly spherical shape  
 (cf. the value of $\mathcal{R} \approx 27$ 
  derived for the outflow of M82 from X-ray spectroscopy in  section~\ref{sec:calc_r_obs_m82}).
Even in the extreme case that the bubbles are compressed into a pancake shape,  
  the effective size of the bubbles relative to the height of the wind cone would not exceed 1/10 (cf. Fig.~\ref{fig:ee00}). 
It is however unlikely that the ionised bubbles would have such an extremely flat pancake-like shape
  as this would require a background flow in the wind cone to have very high speeds,  
  exceeding the thermal velocities of the ionised gas in the bubbles (i.e. ram pressure dominated),  
  yet still maintaining a uniform flow pattern with the absence of turbulence.   
When $\chi = 0$, 
   ${\cal R}$ is independent of $f_2$, the filling factor of the ionised gas bubbles, 
   and hence the fractional volumes of the neutral material and the ionised gas 
   are unconstrained by ${\cal R}$.  
In other words, the total mass loading in the neutral outflow embedded with ionised bubbles 
  cannot be determined using the information obtained from CX line spectroscopy alone.

\subsection{Neutral clumps in an ionised flow} 

\subsubsection{Filling factor}

\begin{figure}
\begin{center} 
\hspace*{-0.0cm}
\includegraphics[width=\columnwidth]{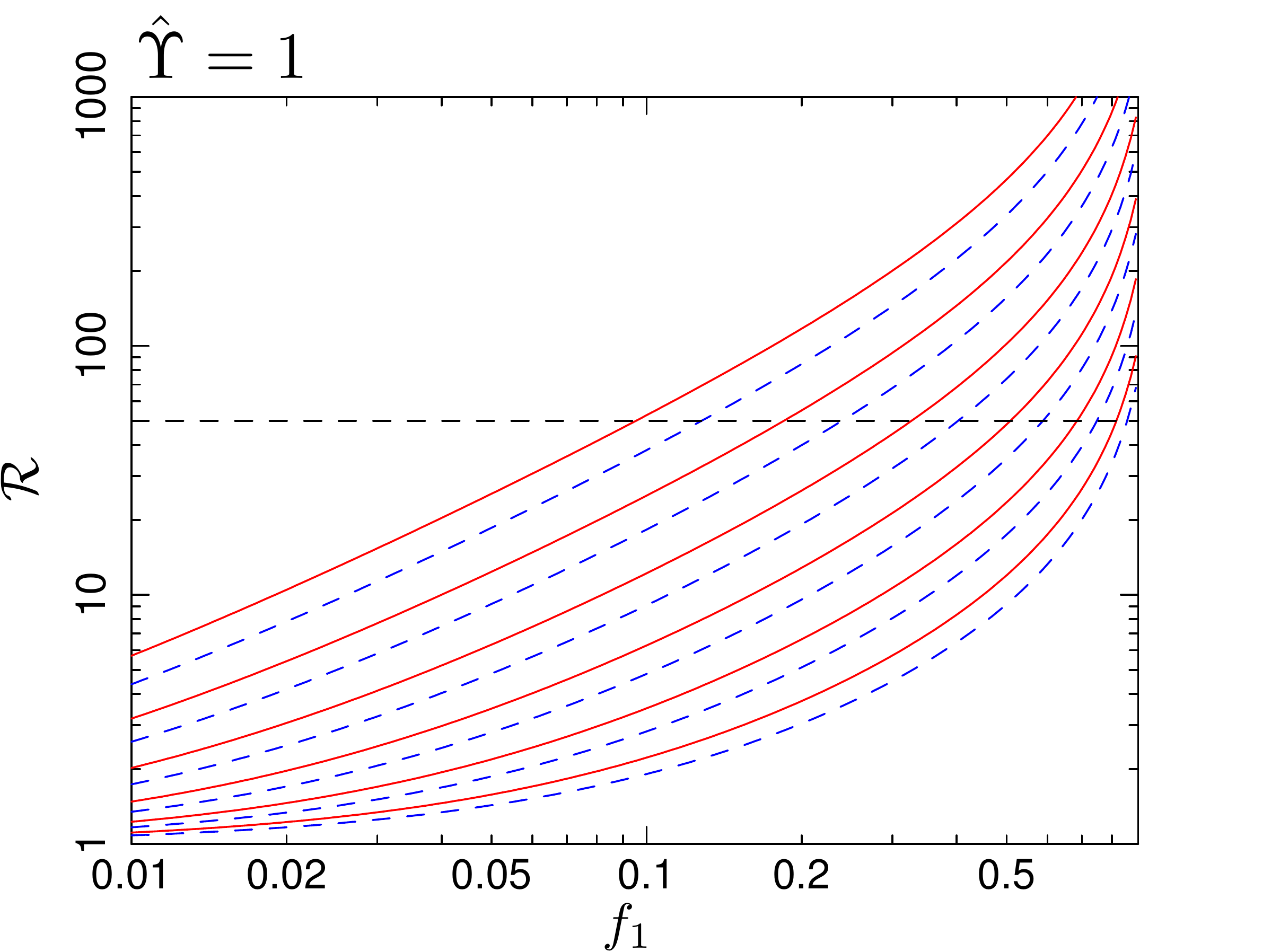} \\
\includegraphics[width=\columnwidth]{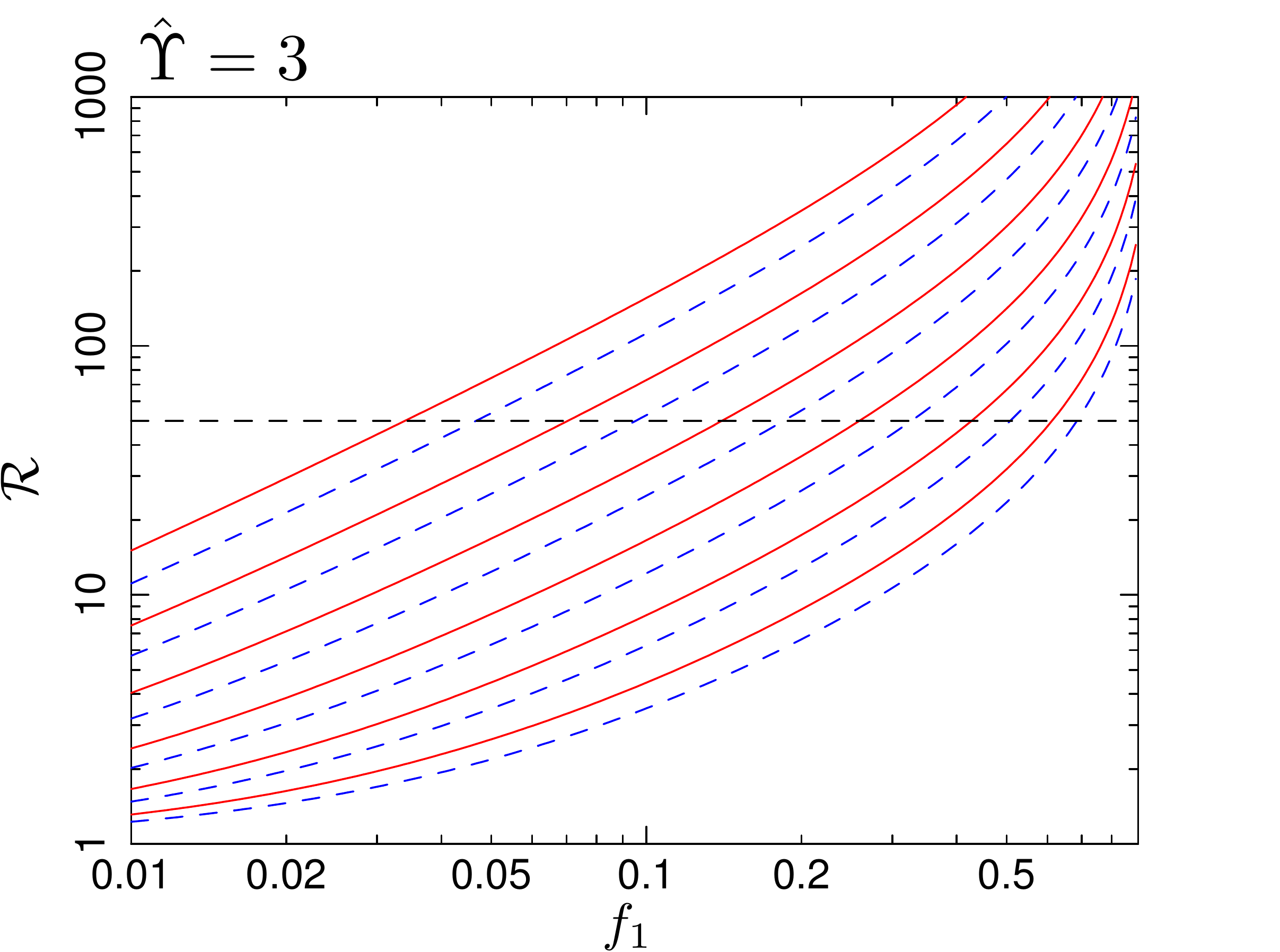} \\
\includegraphics[width=\columnwidth]{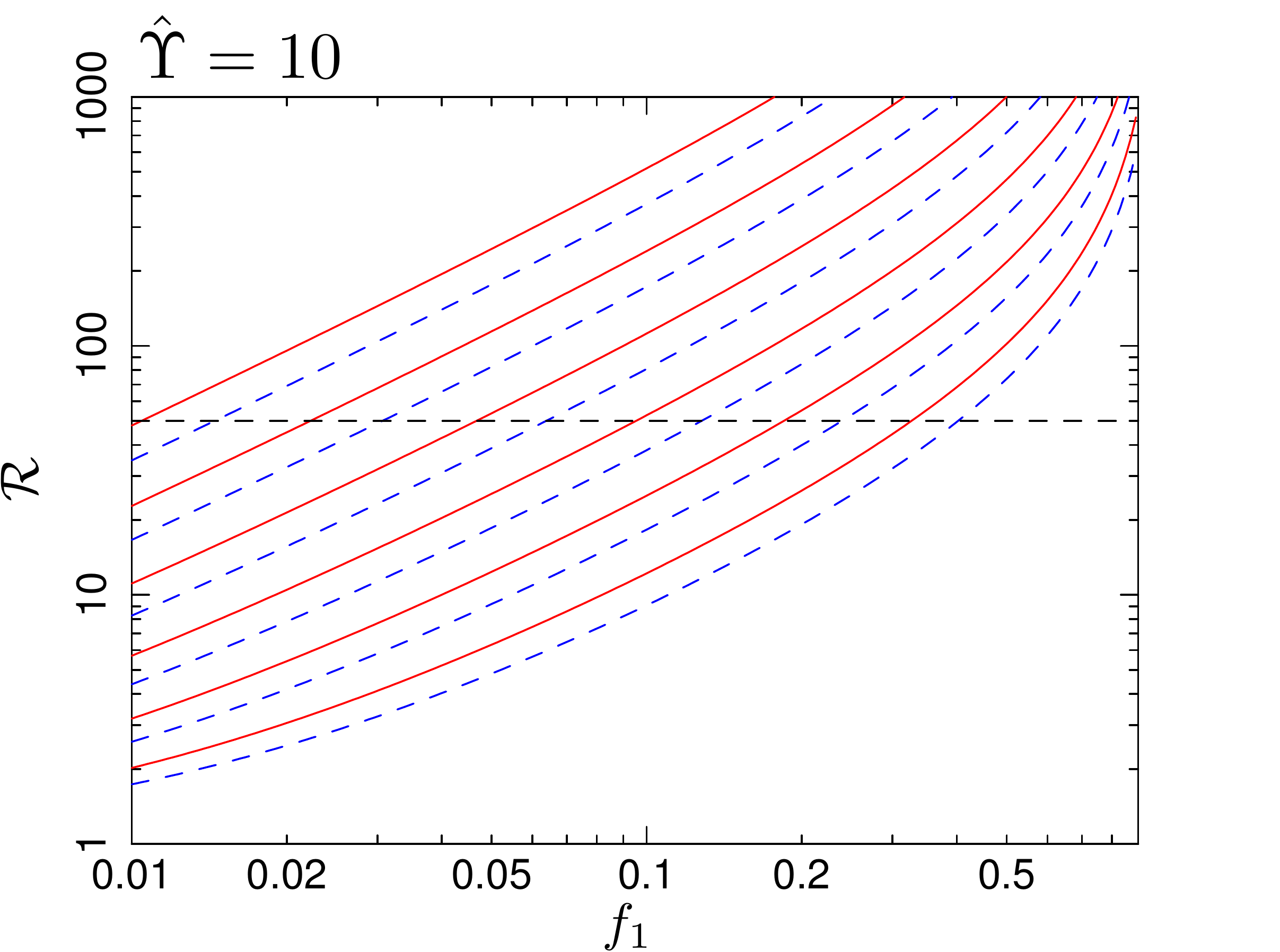} 
\end{center} 
\caption{The relative enhancement of the surface area to volume ratio ${\cal R}$ 
  in the presence of dense neutral clumps entrenched in 
   an ionised outflowing fluid 
   as a function of the filling factor of the clumps $f_1$  
     for ${\hat \Upsilon} =1$ (perfectly spherical clumps), 
     3 (very long filament-like, or moderately flat pancake-like clumps) 
     and 10 (extremely flat pancake-like clumps). 
     Panels from top to bottom.  
  The opening half-angle of the conic wind zone  $\alpha = 60^\circ$ (solid curves) and $30^\circ$ (dashed curves), 
     and the parameter $\chi =1$.   
  In each panel, the set of curves from top to bottom  
     correspond to an increase in the linear size of the dense clumps,  
     with ${\rm Log} (h/r_1) =$  3, 2.6667, 2.3333, 2, 1.6667 and 1.3333 respectively.   }
\label{fig:vpa}
\end{figure}

\begin{figure}
\begin{center} 
\hspace*{-0.0cm}
\includegraphics[width=\columnwidth]{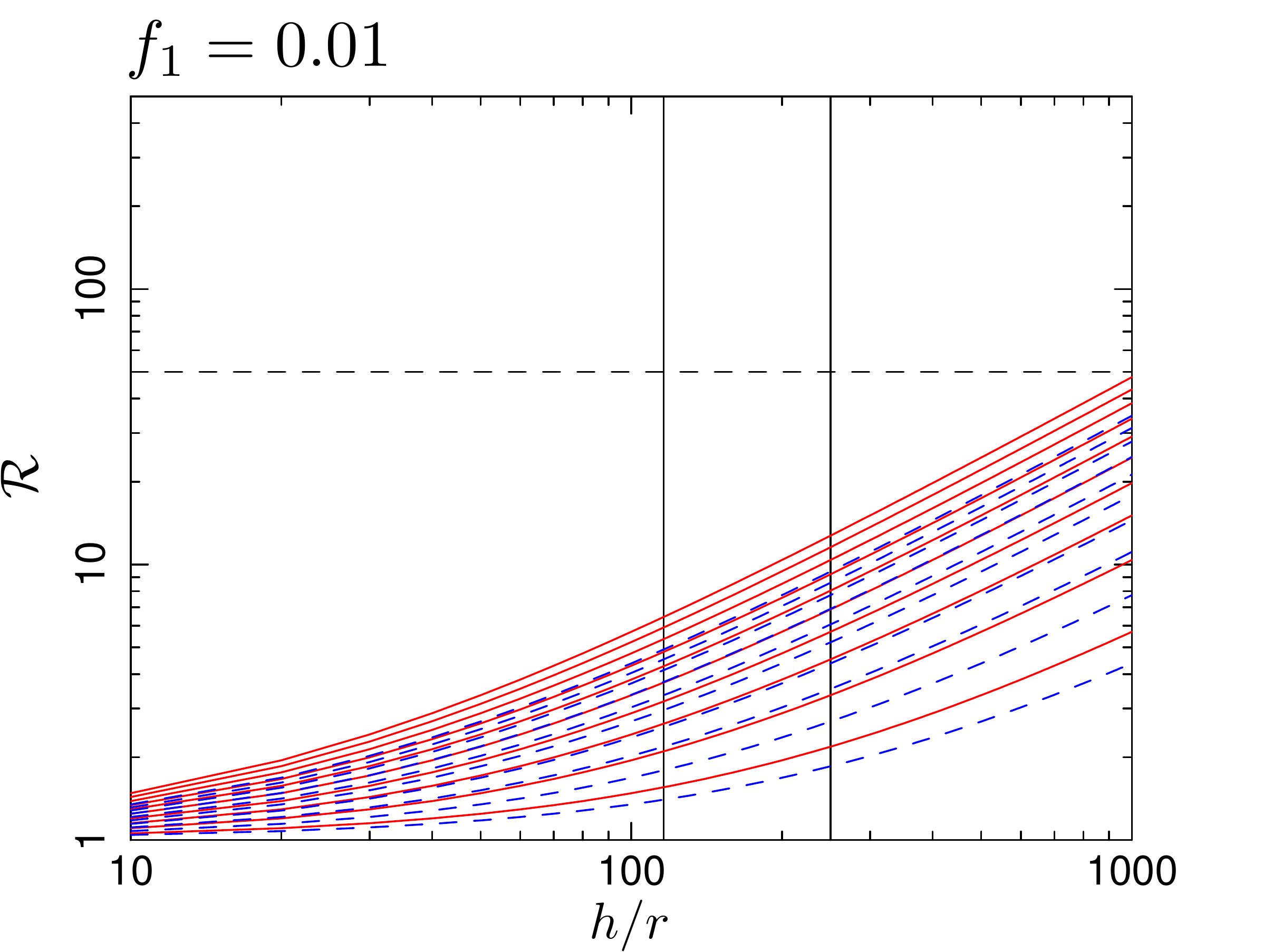} \\
\vspace*{-0.2cm}
\includegraphics[width=\columnwidth]{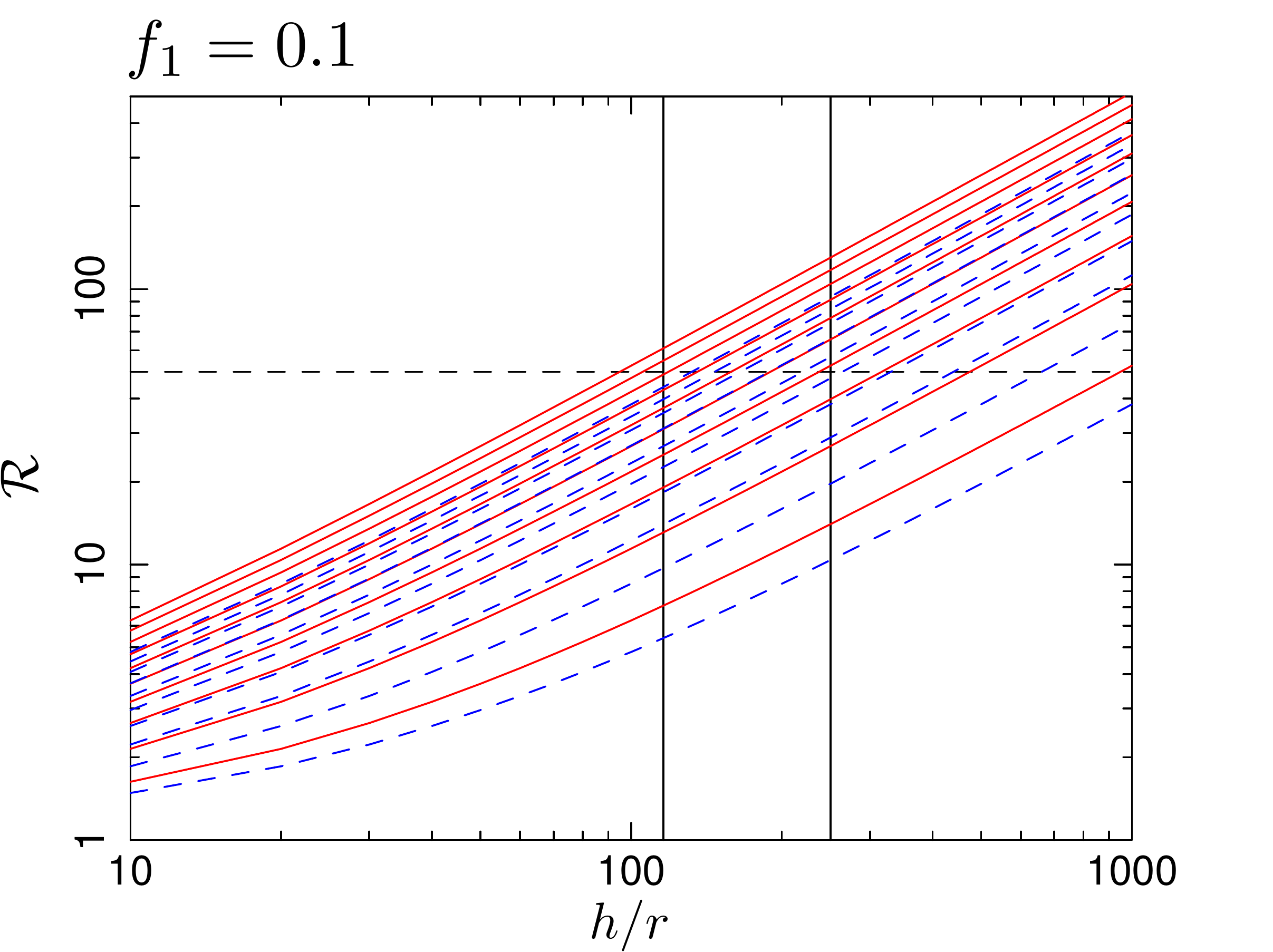} \\
\vspace*{-0.2cm}
\includegraphics[width=\columnwidth]{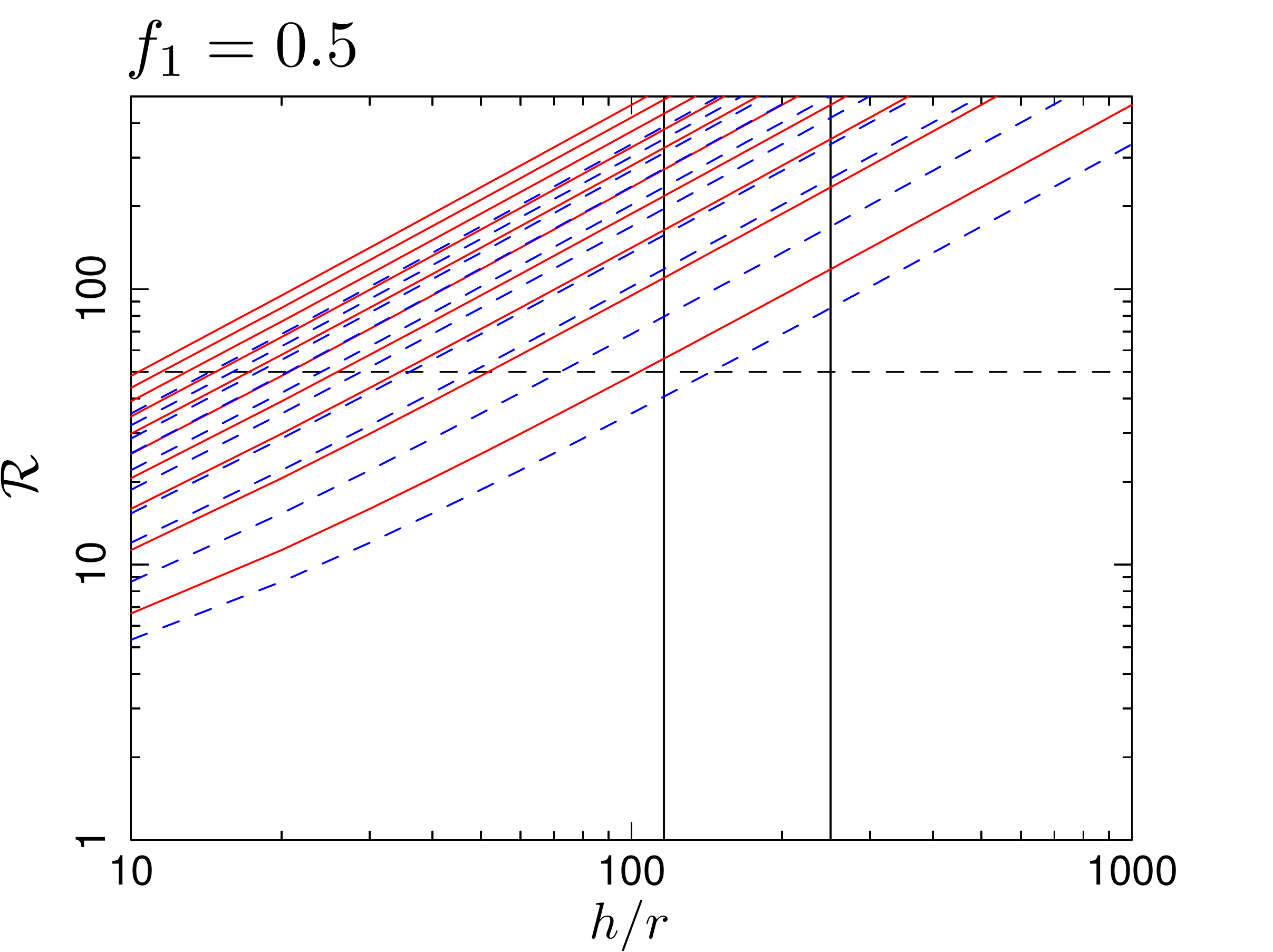} 
\end{center} 
\vspace*{-0.2cm}
\caption{The relative enhancement of the surface area to volume ratio ${\cal R}$ 
   in the presence of dense neutral clumps entrained within 
   an ionised outflow as a function of the parameter $h/r_1$, 
     which specifies the size of the neutral clumps with respect to the height of the wind cone.  
  Panels from top to bottom correspond to the filling factors of the clumps 
     $f_1 = $ 0.01, 0.1 and 0.5.    
  The opening half-angle of the conic wind zone  $\alpha = 60^\circ$ (solid curves) and $30^\circ$ (dashed curves),  
     and the parameter $\chi =1$.   
  The curves from top to bottom in each panel 
     correspond to ${\hat \Upsilon} = 10,$ 9, 8, 7, 6, 5, 4, 3, 2, and 1 respectively.     
  The region below the lowermost curve (with ${\hat \Upsilon} = 1$) in each case is forbidden 
     as no object has a surface area to volume ratio smaller than that of sphere. 
  The two vertical lines represent $h/r_1 = 250$ and $250/(10)^{1/3}$.
  The former corresponds to a spherical clump with a radius $r = 40\;\! {\rm pc}$ (i.e.\ $\hat \Upsilon = 1$) 
    for a wind cone with a height $h = 10~{\rm kpc}$;   
   the latter corresponds to the equivalent radius of a long ellipsoidal-filament with an aspect ratio $\zeta = 10$ 
      and a cross-sectional radius of $40~{\rm pc}$.   }
\label{fig:f00}
\end{figure}

We next consider the case where $\chi =1$,    
  i.e. when the wind cone is one-zone, predominantly filled with ionised gas entrenched with neutral clumps. 
There are two kinds of interfacing surfaces between the ionised gas and the neutral material in the flow 
  where CX emission lines could arise: 
   (i) the boundary surface of the wind cone where the ionised gas meets the surrounding neutral material,  
   and (ii) the surface layer of the neutral clumps carried by the flow. 

By setting $\chi =1$ in equation (\ref{eq:RE}), we obtain      
\begin{equation}  
  {\cal R}  =  \frac{(1+{\rm cosec} \alpha) + \frac{h}{r_1} { {\hat \Upsilon}_1 f_1} }
      {(1+{\rm cosec} \alpha)(1-f_1)} \ .   
\label{eq:RE-c1}
\end{equation} 
The enhancement of the surface area to volume ratio ${\cal R}$ 
  is now determined by the volume filling factor of the neutral clumps 
  as well as their individual sizes and geometries.  
Fig.~\ref{fig:vpa} shows ${\cal R}$ as a function of $f_1$, the filling factor of the neutral material in the flow
  for various effective linear sizes of the clumps $r_1$, expressed in terms of $(h/r_1)$. 
Three clump geometries are shown in the panels from top to bottom: 
  spherical clumps (${\hat \Upsilon} = 1$),  
 filament-like and pancake-like clumps which have large aspect ratios (${\hat \Upsilon} = 3$ and 10 respectively). 
Fig.~\ref{fig:f00} shows ${\cal R}$ as a function of the effective linear clump size, 
  in terms of $(h/r)$, for various ${\hat \Upsilon}$ values. 
The panels from top to bottom correspond respectively 
  to the cases with filling factors of $f_1 = 0.01$, 0.1 and 0.5.   
The general trends shown in Fig.~\ref{fig:vpa} and \ref{fig:f00} are that 
  either smaller effective neutral clump sizes $r_1$, and/or larger ${\hat \Upsilon}$ values for individual clumps, will both give larger values of ${\cal R}$. 
Moreover, ${\cal R}$ increases with the clump filling factor $f_1$.  
An observed value of ${\cal R}$ would therefore yield constraints on the clump filling factor $f_1$
  as well as the clump geometry\footnote{
  We note that, in addition to the large observational and model uncertainties in derived values of $\mathcal{R}$ 
  (c.f. section~\ref{sec:calc_r_obs_m82}), 
  there is some degeneracy between the shape and size of the clumps 
  and the filling factor $f_1$ and their impact on $\mathcal{R}$. 
  Unless these degeneracies can be broken, 
  the constraining power of $\mathcal{R}$ on the internal multi-phase structure on outflows may not be tightened. 
  This issue would be exacerbated by complex (e.g. fractal-like) geometry, which 
  may occur at the interfaces 
  between different phases of material  
  when Kelvin-Helmholtz and/or Rayleigh-Taylor instabilities 
  are present in an outflow.} 
  (i.e. the effective linear size in terms of $(h/r)$ and the aspect ratio as parametrised by $\hat \Upsilon$).  
  
\subsubsection{Characteristic clump size} 
  
In an outflow where the filling fraction of the neutral clumps is non-negligible, 
   the characteristic clump size 
   can be estimated using equation~\ref{eq:r_factor_obs1}, i.e. as:
\begin{equation}
\langle r_1 \rangle \approx \frac{h}{\cal R}\left[ \frac{f_1}{(1+{\rm cosec}\,\alpha)(1-f_1)} \right] 
\label{eq:spherical_r}
\end{equation}   
  where $h$, the height of the outflow cone, and $\alpha$, the opening angle of the cone, can be determined by imaging observations. 
  ${\cal R}$ can be derived from the relative strength of the CX lines. 
The volume fraction of the ionised gas can in principle be obtained from 
  X-ray spectroscopic measurements (assuming a sensible model for the emission process), 
  and hence the volume filling fraction of clumps $f_1$ can be estimated.  
This gives sufficient information for the characteristic size $\langle r_1 \rangle$ of the neutral clumps 
  entrenched within the ionised gas outflow to be constrained.   
If filament-like structures are predominant within the outflow, 
   extreme prolate ellipsoids (or cylinders)
   may be used as a more suitable approximation 
  for their morphology in the context of 
  parametrising their surface to volume ratios.
In this case, the characteristic clump size 
    is modified according to equation~\ref{eq:RE-c1}, thus yielding
\begin{equation}
\langle r_1 \rangle \approx \frac{h {\hat \Upsilon}_1 f_1}{(1+{\rm cosec}\,\alpha) \left[ {\cal R} (1-f_1) -1 \right] } \ ,
\label{eq:spherical_r}
\end{equation}   
  for which the extra parameter ${\hat \Upsilon}_1$ 
  may be estimated from the clump morphology (prolate ellipsoids or cylinders), 
  e.g. from infra-red (IR) emission which more directly reveals the structure of the cold material in the flow.

\subsubsection{Calculating $\mathcal{R}$ for the M82 superwind}
\label{sec:calc_r_obs_m82}

The {\it XMM-Newton} RGS spectrum of the M82 superwind 
  cannot be explained using an optically-thin thermal-plasma model, 
  but a good fit can be found 
  if CX emission is taken into account. 
The CX emission would contribute 
  about one-quarter of the flux in the RGS wavelength range 
  (6$-$30 \AA), 
  and the hot plasma would have a temperature of $\sim$0.6 keV 
  and solar-like metal abundances~\citep{Zhang2014}.
Given that CX process arise at the surface of the neutral material, 
  the observed CX component allow us to determine 
  the effective area of the interface $A_{\rm ion}$ 
  between the hot plasma and the neutral gas.
The normalisation of the ACX model 
  \citep{Smith2012AN, Smith2014ApJ}\footnote{Also detailed online, at~\url{http://www.atomdb.org/CX/}.}, which represents the CX component, suggests a total flux of ions onto the M82 wind's interface area, $S_{\rm int}$, of 
\begin{equation}
\int_{S_{\rm int}} {\rm d}S \   {n_{\rm H}}\;\!  v 
  = 3.2\times10^{51}\,\rm{s^{-1}} \ ,
\end{equation}
with a measurement uncertainty of 15\%, plus an assumed 30\% ACX model uncertainty. Here $n_{\rm H}$ is the hydrogen number-density of 
  the hot plasma, 
  and $v$ is the relative velocity between the hot gas and neutral material.  

Assuming the hot plasma to be homogeneous and has a unique interacting velocity $v$, it follows that $A_{\rm ion}=\int {\rm d}S$.
The density $n_{\rm H}$ of the hot plasma can be inferred from the normalisation of the APEC (Astrophysical Plasma Emission Code;~\citealt{Foster2012ApJ}) model, given by
\begin{equation}
10^{-14}\int_{\rm wc} {{\rm d} V_{\rm ion}}  \   
   \frac{n_{\rm e}n_{\rm H}} 
   {4\pi D^2}=0.0062 \ ,
\end{equation}
with an uncertainty of 20\%. 
Here, "wc" denotes the region over which the integral is performed 
   (i.e. the wind cone), $V_{\rm ion}$ is the volume of the ionised gas in the wind,
  and $n_{\rm e}$ is  the electron density 
  (with $n_{\rm e}\simeq 1.2\;\! n_{\rm H}$  for solar-like abundance).  
We set the distance to M82 $D=3.52\,\rm{Mpc}$ \citep{Tully2009AJ}. 
Without losing generality, 
  we consider that all the ionised gas is contained by 
  the conical outflow,  
  which has a half-cone opening angle $\alpha \approx 30^{\circ}$ 
  and a height $\chi h\sim$3~kpc,  
  and ignore the volume filling fraction 
  of the dense neutral material inside the ionised outflow.  
Then,   
\begin{equation}
V_{\rm ion} = 
\int_{\rm wc} {\rm d}V_{\rm ion}\approx 
  \frac{1}{3}\left(2\pi \chi^3h^3{\rm tan}^2\alpha\right)
  =5.5\times10^{65}\,\rm{cm^3} \ , 
\end{equation}
  and the density of plasma is estimated to be $n_{\rm H}= 0.04\,\rm{cm^{-3}}$. 
While the hot ionised plasma has a velocity $\gtrsim 10^3~\text{km}\;\!\text{s}^{-1}$ 
  in the outflow,  
  the entrained cool gas was found to move substantially slower    
  (at $\sim 500 ~{\rm km}\;\!{\rm s}^{-1}$; see \citealt{Melioli2013}).
An approximation of the relative velocity is 
  then $v\simeq 500~{\rm km}\;\!{\rm s}^{-1}$ $(\pm 30\%)$, 
  which is comparable to the sound speed of the hot plasma.
It follows that an estimate to the total surface area of the interfacing boundaries  
\begin{equation} 
  A_{\rm ion}=\int_{\rm wc} {\rm d}S = 1.6 \times10^{45}\,\rm{cm^2} \  ,
\end{equation} 
  
when integrated across all interfacing surfaces in the wind cone.

Along the axis of the superwind, \ion{H}{i} was detected up to 10 kpc ($h_1$) toward 
  the south and beyond 5~kpc ($h_2$) to the north \citep{Martini2018ApJ}\footnote{There are indications that the northern ionised wind has broken the 3~kpc shell, and could reach as far as the 11.5~kpc `cap'. The wind density is low, and its contribution to the soft X-ray emission is much less than that of the southern wind~\citep{Zhang2014}.}.
Taking this as the entire wind bi-cone, the total surface area of the ionised wind is then 
\begin{equation}
A_{\rm w}=
   \frac{\pi}{3} \left(h_1^2+h_2^2 \right)
    \;\!{\tan}^2\alpha \;\! \left(1+{\rm cosec}\,\alpha\right) 
    =1.25\times10^{45}\;\! {\rm cm}^2 \ ,
\end{equation}
   and the total volume is 
\begin{equation}
V_{\rm w}= \frac{\pi}{3} \left(h_1^3+h_2^3\right)\;\! {\rm tan}^2\alpha =1.15\times10^{67}\,\rm{cm^3} \ .
\end{equation}
The relative enhancement of the area-to-volume ratio then follows as
\begin{equation}
\mathcal{R}=\frac{A_{\rm ion}/V_{\rm ion}}{A_{\rm w}/V_{\rm w}}\simeq 27 \ .
\end{equation}
The uncertainty in this value is dominated 
  by that of the two volumes, 
  $V_{\rm ion}$ and $V_{\rm w}$. 
The value of $\mathcal{R}$ 
  could therefore be off by a factor of a few, 
  given that X-ray emission 
  from the hot ionised gas is dominated 
  by the southern wind cone in M82, 
  which leads to uncertainties in the hot gas density of 50\% and interfacing area of 60\% 
  (neglecting the assumed 30\% ACX model uncertainty).
We also note that 
  the filling factor of the neutral clumps 
  has been ignored 
  in the above estimation, 
  which in term would lead  
  to a value of $\mathcal{R}$ 
  smaller than  
  the expected value 
  when the filling factor of the neutral clumps 
  is considered.

\section{Properties of charge-exchange line emitting clumps}   

\subsection{Stripped gas and H$\alpha$ filaments}
\label{sec:halpha_filaments}

CX processes take place at the boundary between the gases in the neutral and the ionised phases, 
  and therefore CX lines carry information 
   about how the cool neutral gas interacts with the hot ionised gas in galactic outflows.  
Observations have shown H$\alpha$ emitting filamentary structures 
  in the outflows of starburst galaxies  
  (see e.g. the HST H$\alpha$ images of M82 -- \citealt{Mutchler2007PASP}). 
The H$\alpha$ trails are even more apparent in images 
  obtained by the William Herschel Telescope (WHT)\footnote{See \url{http://www.iac.es/telescopes/IAM/2011/73_may11_m82.jpg.html} 
  for WHT RGB + H$\alpha$ image.}.  

Structures are seen   
  in the {\it Chandra} X-ray image of the M82 outflow, at 0.3$-$1.1 keV and 0.7$-$2.2 keV 
  \citep{Kilgard2011AAS}\footnote{The multi-band X-ray images are also available here: \url{http://www.chandra.harvard.edu/photo/2011/m82/}.}. 
`Pancake'-like morphologies 
  were present at high flow altitudes in the 0.7$-$2.2 keV image. 
These X-rays were due, at least in part, 
  to CX processes 
  operating in the interface between ionised and neutral material in the outflow. 
CX lines, including those from \ion{C}{vi}, \ion{N}{vi}, \ion{N}{vii}, \ion{O}{viii} 
  and (most importantly) \ion{O}{vii}, 
  are present in the 0.3$-$1.1 keV band, 
  while \ion{O}{viii} is present also in the 0.7$-$2.2 keV band \citep[see][]{Smith2012AN}.
Although spectroscopic fits to the lines in the RGS spectrum of M82 
  indicate multiple origins for the \ion{O}{vii} triplet, 
  very substantial contributions must come 
  from the CX process.\footnote{
   The line ratios of \ion{O}{vii} triplets were found to be different 
  at different locations in M82. 
  If their physical origins were attributed 
  solely to the same process, we would expect the same line ratios  
  \citep[see observations and discussions in][]{Liu2012MNRAS}. }
\citealt{Liu2011} estimated that CX processes contribute  
  90\% to the \ion{O}{vii} K$\alpha$ triplet lines 
  (with substantial but lower fractions for other species) 
  in the {\it XMM-Newton} RGS spectrum \citep[see also][]{Liu2012MNRAS}.
  
Note that the presence of \ion{O}{i} (63$\mu$m) and \ion{C}{ii} (158$\mu$m) line emission 
  from M82's outflow cones 
  suggests the presence of diffuse \ion{H}{i} gas, i.e. cold neutral gas 
  with temperature below $10^4~K$, 
  while detection of [\ion{O}{iii}] emission 
  (88$\mu$m) 
  would suggest this diffuse \ion{H}{i} 
  is interspersed with \ion{H}{ii} regions \citep{Leroy2015ApJ}, 
  i.e. hot ionised gas with a temperature of $10^6$~K or even above  
  \citep[see][]{Franceschini2000astroph,Contursi2013A&A}. 
Moreover, CO emission is also observed around M82.  
This comes from cold, dense molecular gas 
  \citep[e.g.][]{Franceschini2000astroph, Draine2011book}, so 
presumably, the CO emission is associated with the clumps 
  in which the inner cores are comprised of cold gas 
  harbouring neutral species, 
  or species in low ionisation states. 
This emission is clumpy in morphology 
  and extended along the direction of the outflow cones~\citep{Leroy2015ApJ}. 

Observations of other nearby starburst galaxies 
   also show evidence of multi-phase clumpy outflows. 
For instance, CX X-ray emission 
  (with a strong forbidden line in the \ion{O}{vii} K$\alpha$ triplet) 
  has been detected in NGC~253, M51, M82, M61, NGC~4631 
  and the Antennae galaxies. 
There is also evidence of a combination of thermal and CX X-ray emission 
  in M94 and NGC 2903~\citep{Liu2012MNRAS, Wang2012AN}.

\begin{figure}
\begin{center} 
\hspace*{-0.8cm}
\includegraphics[width=\columnwidth]{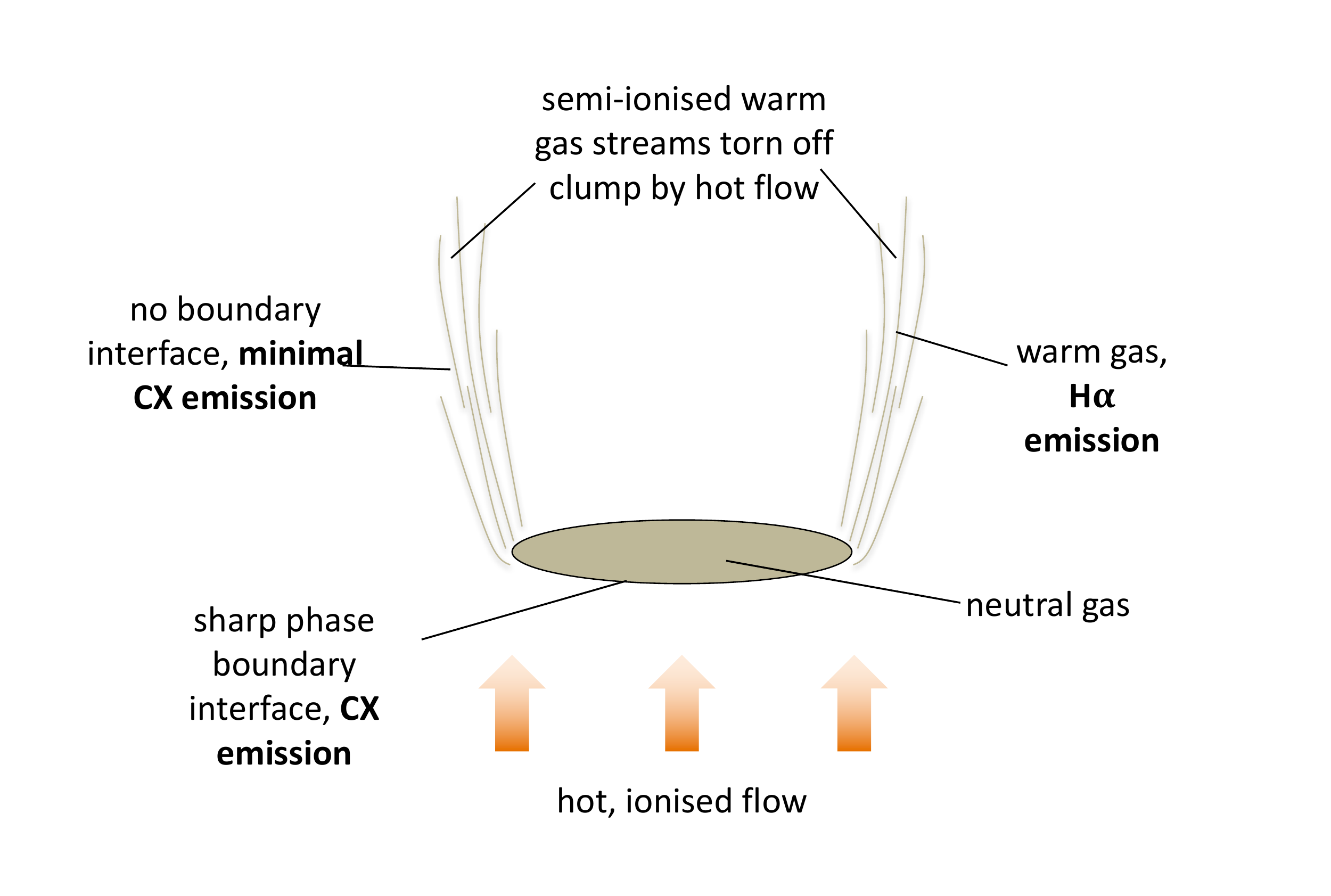}
\end{center} 
\vspace*{-0.25cm}
\caption{Schematic illustration of a cold clump in a hot outflow showing the process by which material is stripped from the surface 
  in contact with the hot outflowing gas.
The sharp boundary between the clump 
  and the surrounding gas is the main source of CX emission.
Trails of semi-ionised warm gas form filamentary structures 
 ahead of the clump, comprising of gas torn off 
 from the clumps by the surrounding flow.
These warm gas filaments emit in H$\alpha$, and are heated by conduction and ionisation as they are exposed to the outflow environment when trailing away from the clump. 
The clump itself is dense and self-shields against much of the external ionising radiation, with heating only operating effectively by conduction over much longer timescales.
       }
    \label{fig:clump_survival}
\end{figure}

This filamentary geometry appears to contradict 
  our deduction (from the CX lines) 
  that the neutral clumps have flattened oblate geometries, 
  i.e. shapes 
  resembling a hamburger, or even a pancake. 
This apparent dilemma can be resolved as follows. 
The H$\alpha$ filaments are not the CX line emitters.  
They are trails of gas torn off from neutral clumps  
  by the fast flowing gas around them   
\citep[see simulations of][]{Suchkov1994ApJ, Cooper2008ApJ, Cooper2009ApJ,
    Scannapieco2015ApJ, BandaBarragan2016MNRAS, 
    Bruggen2016ApJ, BandaBarragan2018MNRAS, Goldsmith2018MNRAS}. 
The stripped material from the cooler CX emitting clumps, 
  when warmed appropriately,  
  emits H$\alpha$ lines and appears as H$\alpha$ filaments. 
Moreover, dense clumps are  
  more likely to survive 
  in the presence of 
  strong irradiation in a galactic outflow environment 
  than geometrically thin filaments 
  -- see Appendix~\ref{A:clumps_vs_filaments}.


\subsection{Survival of clumps}

An approximate pressure balance 
  between the gases of the neutral phase and the ionised phase 
  implies that the neutral clumps, which are cooler, should be denser. 
While the ionised gases are accelerated mechanically or radiatively 
  to form a galactic outflow, 
  the neutral clumps (which have larger inertia) 
  are not accelerated very efficiently.  
A pressure is therefore exerted on the slow-moving clumps 
  by the faster moving ionised gas from below, 
  causing compression of the clump material. 
The shear between the fast-moving ionised gas and the slow-moving clumps 
  would also cause stripping, especially at the edges of the clumps 
   (see the schematic illustration in Fig.~\ref{fig:clump_survival}).
The stripped material naturally has a filamentary structure   
  but is less dense than the cooler clumps. 
The filaments are thus thinner optically as well as geometrically.  

For H$\alpha$ emission to be produced, 
   the material must be warm, with temperatures of around 10$^4$~K or slightly higher,  
   but is not fully ionised. 
Although the filaments are not fully ionised initially, 
  they are in direct contact with the hot ionised gas 
  and fully exposed to the ambient radiation.  
The filaments are therefore heated by particle collisions, radiation from starlight and 
   down scattered X-ray radiation of the ionised gas\footnote{
   The H$\alpha$ emission is unlikely to be caused by the CX X-rays because the ionising (UV/X-ray) continuum due to the hot gas and stellar/SN emission carries much more energy and has more photons than the CX lines.} 
    and also a certain degree of thermal conduction (see, e.g. \citealt{Draine2011book} and also section~\ref{sec:entrained_clumps}). 
In contrast, 
  the material in the inner core of a dense clump 
  is shielded from UV radiation and soft X-rays. 
The timescale for conduction heating is also long 
  compared to the galactic outflow timescale 
  (see section~\ref{sec:entrained_clumps}), 
   so a shielded clump would remain cold and will not be strong in 
   X-ray/UV emission, or even optical. 
Their presence may, however, be tracked by 
  the trails of material that is stripped from them.  
  
A possible origin if these CX emitting clumps 
  is the dense interstellar medium (ISM) of the galaxy 
  entrained into the wind cone 
  and accelerated by the flow  
  \citep[see e.g.][]{Heckman1990ApJS, Cooper2008ApJ, Fujita2009ApJ, Sharma2012ApJ},  
  which pushes 
  on the lower surface of clumps\footnote{For the prescription of an outflow adopted in this model, the dominant pressure acting on these clouds is ram pressure, being around an order of magnitude greater than thermal pressure, with 
  $P_{\rm ram}/P_{\rm th} \sim 10$.}.  
There are several possible consequences:
  (i) the clump is accelerated in the direction of the flow, 
  but it traverses the outflow cone before it is transported to large altitudes; 
  (ii) the clump is entrapped 
  and accelerated sufficiently to become entrained in the plasma fluid; 
  (iii) the clump is evaporated before it can penetrate sufficiently far into the outflow cone for its dynamics to be significantly affected. 
We shall assess each of these possibilities accordingly.

\subsubsection{Acceleration and entrainment} 
\label{sec:acceleation_clumps}

In M82, filaments and clump-like structures move along with an outflow 
  with velocities ${\tilde v}_{\rm clump} \sim 600~\text{km}~\text{s}^{-1}$~\citep[see][]{Strickland2009}.  
The CX line emitters 
  are embedded structures in an outflow 
  and would have similar velocities. 
Thus, we may assume ${v}_{\rm clump} \approx {\tilde v}_{\rm clump}$.
Consider that a clump entering the outflow is accelerated 
  to the observed velocities of those entrained in the flow.
With an initial zero upward velocity, 
  the velocity difference between the clump 
  and the outflowing fluid 
  $\Delta v$ 
  is specified by the outflow velocity alone.  
Without losing generality, 
  we adopt the centre point of the range suggested for M82, 
  i.e. setting $v_{\rm flow} = 1,800~\text{km}~\text{s}^{-1}$.
The fluid plasma outflow velocity is expected to rapidly reach 
  terminal velocity at low altitudes~\citep[see e.g.][]{Chevalier1985}, 
  so the ram pressure acting on an outflow clump
  could be approximated as
  $P_{\rm ram} = \rho_{\rm flow} v_{\rm flow}^2$ throughout the outflow cone.
There would be a decrease in the rate of acceleration  
  as the clumps evaporate 
  (by reducing the cross-section surface), 
  while the velocity offset between the clump and the hot surrounding gas 
  decreases when the clumps are accelerated.   
These complications are not essential for the illustrative estimation here, 
  for which only a first upper-estimate on the timescale is required.    
If assuming a roughly uniform acceleration, 
  then we have    
  $a \sim \sigma_{\rm clump}\;\!P_{\rm ram}/M_{\rm c}$, 
  where $M_{\rm c} \sim 4\pi \;\! n_{\rm H} \;\! m_{\rm H} r'^3/3$ 
  is the mass of the clump, 
  $\sigma_{\rm clump} \sim \pi r'^2$    
  is the effective surface intercepting the hot outflow fluid,
  $n_{\rm H}$ is the hydrogen (number) density of the gas in the clump, 
  $r'$ is the radius of the clump,
  and 
  $m_{\rm H}$ is hydrogen mass. 
This gives a length-scale 
\begin{equation}
\ell_{\rm a} \approx 0.74\;\!\left(\frac{\delta}{10^3}\right) \left( \frac{r'}{10\;\!\text{pc}} \right)  \left(\frac{v_{\rm clump}}{600\;\!\text{km}\;\!\text{s}^{-1}}\right)^{2}\;\!\left(\frac{v_{\rm flow}}{1,800\;\!\text{km}\;\!\text{s}^{-1}}\right)^{-2}  \;\!\text{kpc} \ , 
\end{equation}
   over which the clump is accelerated to 
   velocities comparable to the observed entrained clumps, 
   (with $\delta = n_{\rm H}/n_{\rm flow}$ as the over-density of a clump compared to the density 
   of the outflowing fluid). 
This corresponding acceleration arises on a timescale of
\begin{equation}
t_{\rm a} \approx 2.4\;\!\left(\frac{\delta}{10^3}\right)\;\! \left( \frac{r'}{10\;\!\text{pc}} \right) \;\! \left( \frac{v_{\rm clump}}{600\;\!\text{km}\;\!\text{s}^{-1}} \right)\;\! \left( \frac{v_{\rm flow}}{1,800\;\!\text{km}\;\!\text{s}^{-1}} \right)^{-2} \;\!\text{Myr} \ .
\label{eq:acc_timescale}
\end{equation}

The high-velocity clouds (HVCs) in the Milky Way 
  typically have velocities less than $100~\text{km}~\text{s}^{-1}$ 
  relative to the Galactic standard of rest~\citep[e.g.][]{Wakker1991A&A, Putman2002AJ}.  
If adopting the values observed for the HVCs in the Milky Way  
  as an initial (likely upper bound) velocity of a clump passing into the outflow cone 
  in a direction perpendicular to the flow, 
  we would expect that the HVC would take around 10 Myr to traverse a wind cone 
  of a width of 1~kpc.
This timescale is substantially longer than 
  the expected acceleration timescale (equation~\ref{eq:acc_timescale}).  
The HVCs, and hence the clumps, passing into the flow would therefore
  have sufficient time to be accelerated, 
  and would eventually become entrained into the flow 
  instead of traversing across and leaving it.

\subsubsection{Fate of entrained clumps}
\label{sec:entrained_clumps}

The entrained clumps (and their stripped gas) will be heated 
  by the hot gas in the outflow 
  and also by the ambient radiation, 
  leading to ionisation and evaporation.
For conduction 
 the heat diffusion is expected to be most effective along the minor axis 
  of the clumps (or the filaments),  
  because of both the larger interfacing cross-sectional area 
  and the shorter path from the interfacing surface to the interior core.
CX requires that the neutral material has temperatures $T_{\rm c} \approx 10^2-10^4~\text{K}$ 
  \citep[e.g.][]{Strickland1997A&A, Lehnert1999ApJ}.
The ionised gas in the galactic outflow would have temperatures 
 reaching $T_{\rm h} \approx 10^7~\text{K}$ 
 \citep[see, e.g.][]{McKeith1995A&A, Shopbell1998ApJ}.
The timescale over which a clump (i.e. an entrained HVC) 
  evaporates due to conduction heating is  
\begin{equation}
t_{\rm evap} \approx 1.7\;\!\left(\frac{n_{\rm H}}{10~\text{cm}^{-3}}\right) \left( \frac{r'}{10~\text{pc}} \right)^2 \left(\frac{T_{\rm h}}{10^7~\text{K}}\right)^{-5/2} ~\text{Myr} \ ,
\end{equation}
  \citep[see][]{Draine2011book}, 
  where $n_{\rm H}$ is 
  in the range $10^{-1} - 10^1~\text{cm}^{-3}$ 
  \citep[e.g.][]{Melioli2013}, 
  and  $r'$ is the characteristic clump cross-sectional radius along its minor axis, 
  which is of order 10~pc.

The timescale over which clumps are pushed along the wind cone, 
  can be estimated from their observed velocity and the length-scale of the flow, i.e. 
\begin{equation}
t_{\rm flow} \approx 4.9\;\! \left(\frac{h}{3~\text{kpc}}\right)  \left(\frac{v_{\rm clump}}{600~\text{km s}^{-1}}\right)^{-1} ~\text{Myr} \ . 
\label{eq:t_flow}
\end{equation}
Here, $h\approx3~\text{kpc}$ corresponds to the approximate scale height 
 for the super-wind as those observed in M82~\citep{Strickland2009}   
Note that this parameter varies among systems \citep[see][]{Veilleux2005}.
For our adopted parameter choices, 
  $t_{\rm flow} > t_{\rm evap}$, 
  meaning that clumps (and warm filaments) 
  would generally evaporate and dissolve 
  into the surrounding hot fluid 
  before they can reach the top cap of the wind cone.  
Moreover, comparison with the clump acceleration timescale (equation~\ref{eq:acc_timescale}) 
  indicates  
  that clump entrainment/acceleration would also arise on comparable timescales, 
  i.e. with $t_{\rm a} \sim t_{\rm flow} \sim t_{\rm evap}$. 
Thus, while some clumps entering an outflow 
  are gradually accelerated and evaporated, 
  other dense clumps could survive for their entire passage up the flow,  
  with the neutral gas they harbour being ultimately advected into circumgalactic space.

\subsection{The circumgalactic environment}

\subsubsection{Cooling}

The thermal cooling timescale of ionised gas is   
$t_{\rm cool} \sim n_{\rm e}k_{\rm B}T_{\rm e}/
  \Lambda_{\rm cool}(n_{\rm e},T_{\rm e},Z)$, 
  where $\Lambda_{\rm cool}$ is the cooling rate, 
  $n_{\rm e}$ is the electron number density, 
  $T_{\rm e}$ is the electron temperature, 
  $Z$ is the charge number of ions, 
  and $k_{\rm B}$ is the Boltzmann constant. 
For ionised hydrogen in free-free cooling, 
\begin{equation} 
  t_{\rm cool} 
  \sim  \;\! \left(\frac{n_{\rm e}}{10^{-2} \;\! {\rm cm}^{-3}} \right)^{-1} \left(\frac{T_{\rm e}}{10^7 \;\! {\rm K}} \right)^{1/2} {\rm Gyr} \ .    
\end{equation} 
Thus, $t_{\rm cool} \gg t_{\rm flow}$ 
  (equation \ref{eq:t_flow})   
  for an ionised  outflow 
  with $n_{\rm e}\sim 10^{-2}\;\!{\rm cm}^{-2}$ 
and $T_{\rm e} \sim 10^7\;\!{\rm K}$.
The hot ionised gas will be advected into the CGM, 
  and will eventually cool outside the host galaxy.  
Partially ionised gas 
  of temperatures $T \sim 10^4\;\!{\rm K} - 10^6{\rm K}$ 
  can be cooled more efficiently 
  through bound-free and bound-bound cooling processes 
  \cite[see][]{Sutherland1993ApJS}.  
When the temperature of the outflowing ionised gas 
  falls below $10^6$~K, 
  bound-free and bound-bound processes will become dominant.  
Condensation would occur in the circum-galactic environment  
  \citep[cf.][]{Putman2002AJ, Putman2003ApJ},  
  with the surviving clump remnants as seeds for density growth.
The condensates formed as such 
  when falling back into the host galaxy, 
  would manifest as objects 
  similar to the HVCs 
  that we observe in the Milky Way \citep[][]{Putman2012ARA&A}.   

\subsubsection{Clump infall}

The loss of kinetic energy of CGM clumps 
  causes their infall back into the host galaxy 
  to become part of its ISM. 
The rate at which this happens is determined by a number of processes, 
  in particular, (i) the ram pressure drag experienced by clumps in the CGM, 
    which leads to their energy loss and hence orbital decay,  
    and (ii) the collision and merging of clumps.
The infall timescale is given by    
\begin{align}
   t_{\rm dyn} &\approx  \frac{\pi}{2}\frac{R^{3/2}}{\sqrt{2 G M_{\rm gal}}}     \nonumber \\
   & \approx 27 \;\! \left(\frac{R}{3\;\!\text{kpc}}\right)^{3/2}\;\!\left(\frac{M_{\rm gal}}{10^{10}\;\!\text{M}_{\odot}}\right)^{-1/2}~\text{Myr} \ , 
\end{align}
  where $R$ is the lengthscale of the CGM and 
  $M_{\rm gal}$ is the dynamical mass of the system,
  which is scaled here to that of M82 
  \citep[see e.g.][]{Strickland2009, Greco2012ApJ}.
The kinetic energy loss of a clump due to ram pressure drag 
  will lead to a decay of its orbit. 
Assuming a uniform deceleration, we obtain a drag timescale of  
\begin{align}
t_{\rm drag} & = \frac{M_{\rm c}}{  \pi \langle r' \rangle^2 \;\! m_{\rm p}\;\!n_{\rm bg} \;\! v_{\rm c}} \nonumber\\
& = \frac{4}{3} \frac{\langle r' \rangle}{v_{\rm c}} \delta \nonumber \\
& = 1.31 \;\! \left(\frac{\langle r' \rangle}{100~\text{pc}}\right) \;\! \left(\frac{v_{\rm c}}{100~\text{km}\;\!~\text{s}^{-1}}\right)^{-1}\;\!
  \left(\frac{\delta}{1000}\right) ~\text{Gyr} \  
\end{align}
   \citep{Maller2004MNRAS}, 
where $\langle r' \rangle$ is the mean radius of the clump 
   (rather than the size of the minor axis), 
   for which we adopt a characteristic value of $100~\text{pc}$~\citep{Crighton2015MNRAS}. 
Here $M_{\rm c}$ retains its earlier definition as the clump mass, 
   $v_{\rm c}$ is its velocity in the CGM, 
   $n_{\rm bg}$ is the number density of the hot component of the CGM gas, 
  and $\delta = n_{\rm H}/n_{\rm bg}$ is the over-density of a clump on the CGM background.
The CGM gas is slightly cooler than that of the outflow fluid, 
   and its temperature is expected to be 
   $T\approx 10^4 - 10^{5.5}~\text{K}$~\citep{Narayanan2010ApJ, Werk2013ApJS, Stocke2013ApJ, Tumlinson2017ARA&A}. 
Thus, the ram pressure drag-induced kinetic energy loss  
  is insufficient to cause galactic outflow material to return  
  and replenish the ISM of the host galaxy.

An alternative mechanism is clump-clump collisions.  
The process is stochastic, in contrast to the drag process
  which extracts a clump's kinetic energy continuously. 
In a collision, some fraction of the kinetic energy of the colliding clumps 
  will be transferred into turbulence, 
  causing heating in the clump material. 
When the thermal energy is radiated away, 
  it leaves a single massive remnant  
  with a kinetic energy lower than those of its pre-merge progenitors~\citep{Maller2004MNRAS}. 
The efficiency of the clump-clump collision process may be estimated as follows. 
The initial number of clumps (per CGM unit volume) is given by 
\begin{align}
N_{\rm c} &\approx \frac{3\;\! M_{\rm neut}}{4 \pi M_{\rm c} R^3} \nonumber \\
& \approx 27 \;\! \left(\frac{M_{\rm c}}{10^6~\text{M}_{\odot}}\right)^{-1} \;\! \left(\frac{M_{\rm neut}}{3\times 10^9~\text{M}_{\odot}}\right) \;\! \left(\frac{R}{3~\text{kpc}}\right)^{-3}~\text{kpc}^{-3} \ , 
\end{align} 
  where $M_{\rm neut}$ is total mass content in the neutral material in the CGM,  
  and $M_{\rm c}$ is the averaged mass of a single clump. 
It is argued that 20-30\% of the dynamical mass of the host galaxy  
  may reside in the cold, neutral component of the CGM~\citep[e.g.][]{Maller2004MNRAS, Stocke2013ApJ}.
The neutral material is likely to be condensates, i.e. in the form of clumps, 
  rather than being mixed-in with the ionised material\footnote{Note that here the neutral clumps are referred to in a general context, i.e. in that they are dense condensates in the CGM. 
They can be formed by condensation of the CGM directly, or by condensation of the CGM 
    seeded by the remnant clumps which have survived their passage through the galactic outflow to be advected into the circumgalactic environment,  
    or by the recycling condensation of the galactic outflow material.}. 
For a galaxy similar to M82, 
  the amount of neutral gas in the CGM may therefore be 
  $M_{\rm neut} = 3\times 10^{9}~\text{M}_{\odot}$ 
  \citep[see e.g.][]{Maller2004MNRAS, Stocke2013ApJ, Werk2013ApJS}.  
Presumably, the end-state of CGM clumps after they have undergone an infall 
  would be HVCs. 
The masses of the Galactic HVCs are found to be within a range of 
  $10^5 - 5\times 10^6~\text{M}_{\odot}$~\citep[e.g.][]{Putman2012ARA&A}.  
Without losing generality, 
  we assign $M_{\rm c} = 10^{6}~\text{M}_{\odot}$ 
  as the characteristic mass of CGM neutral clumps. 
(We acknowledge these are observed to be comprised 
  of smaller sub-structures of filaments and clouds, c.f.
  \citealt{Putman2003ApJ, Thom2008ApJ, Hsu2011AJ, Fox2014ApJ}, also seen in M31 -- \citealt{Thilker2004ApJ}.)

The clump-clump collisional timescale may be estimated as     
\begin{align}
t_{\rm cc} &\approx \left[N_{\rm c} \;\! \sigma_{\rm c} v_{\rm c}\right]^{-1} \nonumber \\
   &\approx 11.8 \;\! \left(\frac{N_{\rm c}}{26.5~\text{kpc}^{-3}}\right)^{-1}\;\!
   \left(\frac{v_{\rm c}}{100~\text{km}~\text{s}^{-1}}\right)^{-1}\;\!
   \left(\frac{\langle r' \rangle}{100~\text{pc}}\right)^{-2}~\text{Myr} \ ,  
\end{align}
  which is shorter than the timescale for drag-induced energy loss 
  for the parameters considered in this work. 
Clump-clump collisions are therefore a viable mechanism 
  for returning the outflow material.  
Note that the collisional timescale estimated above  is slightly longer than 
  the timescale over which clumps are advected 
  (cf. equation~\ref{eq:t_flow}). 
An advection timescale shorter than the kinetic energy loss timescale 
  implies a gradual building up of CGM.   
The accumulation of outflow material 
  would persist until the termination of the starburst phase 
  (hence, the cut off of the energy supply for outflow -- 
  see e.g. ~\citealt{Chevalier1985} for a generic scenario and the hydrodynamics formulation). 
If the star formation occurs in cycles, 
 this ejection of multi-phase clumpy ISM matter will give rise 
 to complexity in the feedback interplay.  
On the one hand, the depletion of cold gas in the ISM 
  and the advection of neutral material to the outside of the host galaxy 
  would undoubtedly suffocate the star formation process.   
  But, on other hand, a multi-phase outflow would provide metal-enriched material, 
  which can be cooled more efficiently than the metal-poor pristine cosmological gas.  
The remnant clumps would be seeds that nucleate CGM condensation, 
  and the subsequent infall of the newly formed CGM condensate 
  would reignite and fuel the next phase of the star formation cycle.

\subsection{Additional remarks} 

The large-scale galactic outflow in M82 
  is believed to be driven by a combination of outward thermal pressure 
  and the coalescence of shocks 
  and outflowing material, 
  with power derived from an active star-forming core 
  at the base of the wind cone. 
In such a setting, 
  the hot outflow would entrain cold neutral material from the ISM 
  which would form into clumps and filaments. 
In a galactic outflow with a single phase medium, 
  an asymptotic flow velocity will develop 
  when the energy injection by the supernova explosions 
  is deposited into the thermal and kinetic energy of the fluid 
  \citep[see][]{Chevalier1985}. 
In a multi-phase clumpy galactic outflow, 
  a fraction of the supernova power 
  will be dissipated for the compression of the cold neutral clumps that are entrained into the hot galactic outflow medium. 
Such compression has already been seen in numerical simulations  
  and is also verified 
  by the flattened hamburger/pancake shapes 
  that we infer from the surface-to-volume analyses 
  using observations of CX emission. 
Another fraction of the supernova power  
  will be dissipated in the drag 
  between the clumps and the gas in the flow, 
  by which the clumps are accelerated 
  \citep[see][]{Larson1974MNRAS, Nath1997MNRAS, Veilleux2005, Bustard2016ApJ}.   
  
A single fluid formulation is therefore inadequate 
  to model the thermodynamics and hydrodynamics of 
  such multi-phase clumpy galactic outflows.  
  It is needless to say that further complications would also arise  
  due to the radiative cooling of the hot outflow medium 
  and the additional pressure exerted 
  by radiation and cosmic rays produced in star-forming regions 
  at the base of the wind cone.  
Moreover,  
  the entrained clumps and filaments are not stationary structures 
  as they are subjected to mechanical ablation, i.e. stripping, 
  by the faster-moving fluid in the surrounding  
  and thermal evaporation by UV and X-rays 
  permeating through the hot galactic outflow medium. 
In addition, 
  clump destruction can also arise due to thermal conduction 
  (see the discussions 
  in sections~\ref{sec:halpha_filaments} and~\ref{sec:entrained_clumps}). 
The material evaporated or mechanically stripped from the clumps 
  will be pushed ahead in the flow.  
This is heated and stretched, 
  forming a semi-ionised H$\alpha$-emitting trail. 
The galactic outflow is likely to be turbulent, 
  and this would cause the filamentary trails to develop 
  fractal-like structures 
  rather than a well-defined interface surface 
  on which CX process would operate in a quasi-steady manner. 
  
The survival of the stripped material against mixing 
  when it extends into the hot galactic outflow  
  depends on many factors and their interplay. 
The dissolution of these fractal trails in a hot outflow 
  is an important avenue of future research, 
  given that it will strongly impact 
  on the mass-loading of the flow, 
  the redistribution of material in different fluid phases, 
  the evolution of substructures, 
  and the observational morphologies of 
  X-ray and H$\alpha$ emission features. 
  
Galactic outflows in active star-forming galaxies 
  do not necessarily 
  take the form of a super-wind, 
   where dense clumps 
   and H$\alpha$ emitting filaments are entrained within,  
   as that of M82~\citep{Chevalier1985, Ohyama2002}. 
In NGC~3077, 
  a starburst dwarf neighbour of M82, 
  bubbles and expanding shells 
  (instead of elongated filaments)  
  are observed ~\citep{Ott2003}. 
The bubbles consist of hot gas,  
 and they are enclosed by a warm shell, 
 characterised by the H$\alpha$ emission. 
In terms of the two-zone model 
 shown in Figure~\ref{fig:wind_cone}, 
 the super-wind galactic outflows 
 would be dominated by the bottom zone, 
 while in the galactic outflows of NGC~3077
 the bottom zone is negligible or absent. 
Such different multi-phase structures  
  would have different hydrodynamic properties. 
These would have impacts on the entrainment of HVCs, 
  if it happens,  
  and the subsequent evolution of HVCs/clumps in the outflow.
 
In fact, 
  the diversities of galactic outflow conditions 
  are now recognised, 
  and outflows from star-forming galaxies 
  can be powered by various driving mechanisms. 
A galactic outflow can be driven by the radiation from 
  an intense star-forming core in the host galaxy~\citep{Dijkstra2008MNRAS, Nath2009MNRAS, Thompson2015MNRAS}. 
It has also been suggested 
  that cosmic rays can play a very important role 
  in regulating the energy budget of galactic outflows 
   ~\citep{Samui2010MNRAS, Uhlig2012MNRAS}\footnote{At high redshifts, the favoured mechanism for driving galactic outflows 
   is cosmic-ray pressure. 
   Studies have argued 
   that cosmic-ray driven outflows become increasingly important 
   in earlier epochs~\citep[e.g.][]{Samui2010MNRAS}, 
   when host galaxy masses were presumably smaller 
   and star-formation rates were higher during bursts.}.
The ability for a neutral clump to survive 
  in a hot ionised galactic fluid, 
  an intense radiative field 
  or a bath of cosmic rays 
  may depend on the metalicity and even the magnetic field 
  in addition to the density of the clump,  
  the outflow speed 
  of the hot ionised gas  
  and the length-scale of the outflow. 

Ignoring magnetism for the time being, 
  in a low-temperature outflow  
  (e.g. in a radiatively driven system, 
  see \citealt{Zhang2018Gal}) 
  clump destruction is less likely to arise 
  by thermal conduction. 
Instead, the abundant ionising radiation responsible 
  for powering the outflow 
  will cause surface evaporation of a clump, 
  particularly 
  on the surface facing the host galaxy, 
  where the strong radiation is emitted 
  from the regions with intense star formation.  
 Outflows driven predominantly by cosmic ray pressure 
   may also exhibit a relatively low fluid temperature 
   in the outflow material 
   \citep[see e.g.][]{Samui2010MNRAS}. 
As the energy deposition of the cosmic rays 
  will not be concentrated at the base of the outflow cone, 
  the gas would accelerate 
  not only by a thermal pressure gradient 
  but also by a cosmic ray pressure gradient, 
  which shares some similarities 
  to the radiative pressure. 
A consequence is that 
  outflows can extend to altitudes higher than 
  those driven purely by supernova power mechanically, reaching a few tens of kpc instead of just a few kpc 
  \citep{Naab2017ARA&A, Jacob2018MNRAS, Girichidis2018MNRAS}.
Again, the lower temperatures 
  reduce the amount of heat transported into clumps 
  by means of thermal conduction, 
  which reduces this as a means of dissolving the clumps 
  and erasing the temperature inhomogeneities within the flow.
 However, the lower velocities sometimes attributed to cold cosmic ray driven outflows~\citep[e.g.][]{Samui2010MNRAS}   
   mean that entrained clumps remain in the (much larger) outflow cone for greatly extended periods of time. 
 This would mean that a given clump/over-density 
   is exposed to eroding and irradiating processes 
   for much longer than its counterparts 
   in a supernova-driven hot outflow, 
   which may still lead to its eventual destruction 
   well before it has completed its passage of an outflow cone.  
   As such, it would not be advected into the circumgalactic space, despite the less efficient heat conduction process. 
 Cosmic rays can cause heating and ionisation 
  in the CGM and the intergalactic medium as well as the ISM 
  \citep[see e.g.][]{Owen2018MNRAS}, 
and tend to deposit their energy into dense targets. 
  Hence the dense cores of entrained neutral clumps 
  are more effective in capturing and absorbing the cosmic ray particles 
  then the galactic outflow medium. 
The heating effect due to cosmic rays 
  can also be enhanced if clumps are magnetised. 
As the energy is deposited mainly in the densest parts of clumps 
  and/or the most strongly magnetised regions,  
  clumps in a cosmic ray dominated flow are being 
  pushed along by their cores 
  (i.e. expanding `inside-out' instead of compression by an external pressure force). 
This could stretch clumps out over time, 
  if the magnetic fields are also stretched 
  along the flow of ionised fluid.  
This will lead to the formation of extended filaments 
  oriented along the outflow direction 
  -- in contrast to the clumps in a supernova-driven system, 
  where pancake or hamburger shape clumps 
  with their flattened surface oriented perpendicularly 
  to the flow direction of a hot galactic outflow fluid.    
The action of different clump-flow interactions 
  should give different surface-to-volume ratios  
  and the interfacing surface would be accordingly 
  marked by the strength of the CX lines in their emission spectra. 
 
Finally, we note that 
  in outflows 
  with a significant magnetic energy density and turbulence, 
  the heating/evaporation of entrained clumps in a hot outflow 
  may be reduced (if cosmic rays are insignificant). 
This is because the thermal particles (usually electrons) 
  responsible for the conduction of heat propagate predominantly along magnetic field lines. 
In the presence of tangled and turbulent magnetic fields, 
  such lines increase the path that particles must travel when conducting heat, 
  thus slowing the conduction process down 
  \citep{Tao1995MNRAS, Chandran1998, Malyshkin2001ApJ}. 
This would operate to protect entrained clumps, 
  enabling them to survive much longer in the outflow cone, 
  and allowing them to persist to higher altitudes.

\section{Summary and conclusions}
\label{sec:summary_and_conclusion}

Galactic outflows are complex multi-phase fluids, 
  with cold dense clumps and warm filaments entrained in a hot ionised gas.  
CX (charge-exchange) processes are a characteristic of fluids  
  consisting of a neutral and an ionised phase. 
For galactic outflows, 
  they operate in the interface surface between 
  neutral or partially ionised clumps and the ionised gas, 
  and the CX processes give rise to X-ray CX emission lines.    
The large-scale outflows of a number of nearby starburst galaxies 
  show a strong forbidden line in the X-ray \ion{O}{vii} triplet \citep{Liu2012MNRAS, Wang2012AN}, 
  which are interpreted as CX lines, 
  thus establishing the multi-phase nature of galactic outflows. 

In this work, we conducted analyses of the surface-to-volume ratio 
  of dense clumps   
  based on observations of CX emission from starburst galaxies, 
  and hence constrained the geometries and structures of the dense neutral material 
  in galactic outflows. 
More specifically we considered the relative enhancement 
  of the area-to-volume ratio of the dense neutral clumps with respect to spheres, 
  and we derived the aspect ratios of these clumps 
  from the observed strengths of CX lines, e.g. in M82~\citep{Liu2011}. 
Our analyses indicated that the cold dense clumps in galactic outflows 
  such as those of M82 
  would have flattened shapes, resembling a hamburger or a pancake, 
  instead of elongated shapes. 
The flattened geometry is consistent 
  with the findings of numerical simulations, 
  which show that dense clumps entrained in galactic outflows 
  would be ram pressure compressed. 
Our analyses do not support an elongated geometry for the CX emission objects. 
Thus, the filamentary features observed in the H$\alpha$ images of galactic outflows  
  are not primary CX emitters. 
We interpret them as 
  warm trails of the stripped material 
  from the colder dense CX emitting clumps, 
  which are pushed forward (after having been stripped) and stretched into filamentary structures   
  by the faster-moving galactic outflow fluid.  
These filaments are exposed 
  to the hot gas in their surroundings 
  and intense background radiation. 
Though cold initially, they are gradually warmed. 
This is consistent with numerical studies 
  which have shown 
  that 
  dense, slowly moving clumps are ablated 
  by shear and ram pressure 
  exerted by the fast-moving galactic outflow fluid, 
  and the stripped material forms long trails leading the clump of their origin.  

We have found that 
  the entrainment, acceleration and evaporation/dissolution of neutral clumps 
  in a galactic outflow occur over comparable timescales 
  (see section~\ref{sec:entrained_clumps}). 
Thus, some fraction of the clumps, and perhaps the stripped gas filaments,  
  can survive their entire passage through the galactic outflow 
  to be advected into the circumgalactic space surrounding their galaxy of origin. 
These remnants are metal-enriched,  
  and they are seeds for the condensation of CGM (circumgalactic medium)  
  to form a new generation of clumps. 
Clump-clump collisions in circumgalactic environments 
  would cause clump infall into the ISM of their host galaxy, as the clumps would lose a substantial fraction of their kinetic energies in a collision.
These new clumps could be the origin of the HVCs (high-velocity clouds) 
  as those observed in our Galaxy~\citep[see][]{Putman2012ARA&A}. 
When infallen clumps are re-entrained into the galactic outflow, 
  a recycling of galactic material is initiated.

\section*{Acknowledgements} 
We thank the referee 
  for helpful comments 
  and suggestions to improve the manuscript.  
KJL's research at UCL-MSSL 
  was supported by UCL through a MAPS Dean's Research Studentship 
  and by CUHK through a CN Yang Scholarship, a University Exchange Studentship, 
  a Physics Department SURE Studentship, and a New Asia College Scholarship.   
ERO was supported by a UK Science and Technology Facilities Council PhD studentship.  
This research has made use of the SAO/NASA ADS system and the arXiv system.  


\bibliographystyle{mnras}
\bibliography{reference}

\appendix

\section{Surfaces areas of ellipsoids} 
\label{A:ep}

The surface area of a triaxial ellipsoid specified by the the semi-axes $a$, $b$ and $c$ 
  (with $a \geq b \geq c$) is given by 
\begin{equation} 
  A = 2\pi c^2 + \frac{2\pi ab}{\sin \phi}
     \left[ F(\phi, k) \cos^2\! \phi + E(\phi, k) \sin^2\!\phi   \right] \ ,  
\end{equation} 
   where $\phi = \cos^{-1}(c/a)$, $k^2 = (a^2/b^2)[(b^2- c^2)/(a^2 -c^2)]$.  
The incomplete elliptic integrals (of the first and second kind respectively) above 
  are defined as 
\begin{equation}  
\left\{ 
\begin{split}  
 F(\phi,k) & = \int_0^\phi {\rm d}\theta \, (1- k^2 \sin^2\! \theta)^{-1/2} \ ;  \\ 
 E(\phi, k) & = \int_0^\phi {\rm d}\theta \, (1- k^2 \sin^2\! \theta)^{+1/2} \ ,  
\end{split} 
 \right.
\end{equation}     
  and they do not have a closed form that can be expressed in terms of elementary functions. 
However, it was shown that the surface area $A$ is left- and right-bounded, with   
\begin{equation}  
  \frac{1}{3} \big(ab +bc+ ca\big) \leq \frac{A}{4\pi} \leq 
  \left[ \frac{1}{3}\left( a^2b^2+b^2c^2+c^2a^2 \right) \right]^{1/2} \   
\end{equation}  
  \citep{Klamkin1971}.  
An approximate formula  
\begin{equation} 
   A = 4\pi \left[ \frac{1}{3} (a^p b^p + b^p c^p + c^p a^p)\right]^{1/p}   
\end{equation} 
  with $p = 1.6075$ was therefore proposed by Knud Thomsen,   
  and it can achieve a very high accuracy (less than 1.061{\%} error).\footnote{See 
    \url{http://www.numericana.com/answer/ellipsoid.htm\#thomsen} for detailed discussions  
    on the subject; see also \cite{Lehner1950,Klamkin1971,Dunkl1994a,Dunkl1994b} and \cite{Rivin2004} 
    for evaluating/computing the surface areas of 3D and nD ellipsoids.} 
  
\section{Surface area to volume ratios of convex objects with aspect ratios close to unity} 
\label{A-one}

Under an approximately constant gravitational field\footnote{ 
  The approximation is valid when $s  \ll  | \;\!\Phi / \nabla \Phi \;\!\! |$, 
  where $s$ is the linear size of the fluid droplets and $\Phi$ 
   is the local gravitational potential.},  
   droplets with a relative speed with respect to a background fluid flow  
   could develop into shapes resembling a hamburger. 
   An example of such is the shape of raindrops in the terrestrial atmosphere.  
In the absence of electric stress and internal fluid circulation, 
   the three key factors that determine the shapes of the droplets  
   are aerodynamic pressure, hydrostatic pressure and (effective) surface tension 
   \citep{Beard1987}.   
Thus, if the clumps and bubbles have a non-negligible surface tension, 
  e.g. such as that provided by some locally divergence-free internal or external magnetic fields 
  at the interfacing surface, 
  they would develop shapes that resemble those of the raindrops. 
As an approximation, we may treat a hamburger shape as a hemisphere 
  when calculating the surface area.  
  We can then obtain an analytical expression 
  for the enhancement of the surface-area to volume ratio with respect to a sphere with the same volume 
  of these raindrop-like dense neutral clumps or ionised gas bubbles:   
\begin{equation}  
  {\hat \Upsilon} = \frac{3}{2^{4/3}} \approx 1.19\ , 
\end{equation}  
  which implies 
\begin{equation} 
  \frac{h}{r} {\hat \Upsilon} = \left(\frac{3}{2}\right) \frac{h}{r'} 
\end{equation} 
where $r'$ is the hemisphere radius and $r$ is the equivalent radius of a sphere. 
In fact, any convex polyhedrons with aspect ratios $\approx 1$,   
   will give $ {\hat \Upsilon} \sim (1 - 2)$. 
Thus, if dense neutral clumps and ionised gas bubbles  
  have aspect ratios $\approx 1$ and,  
  if their surfaces have positive curvatures everywhere, 
  then ${\hat \Upsilon} \sim {\cal O}(1)$. 
  This implies that clumps and bubbles with these geometrical properties 
  cannot give a surface area to volume ratio
  significantly larger than a simple sphere.  
In other words, in order to have large surface-area to volume ratios,   
  the clumps or bubbles must have large aspect ratios, 
  or their surfaces interfacing with the external medium 
  cannot have positive curvatures everywhere.  


\section{Dense clumps and H$\alpha$ filaments}
\label{A:clumps_vs_filaments}

Consider a cold gas cloud consisting of mainly hydrogen and a small amount of ``metal''. 
Let $n_{\rm H}$ be the hydrogen number density, $\varsigma$ be the number ratio of ``metal'' atoms and hydrogen atoms, 
  and $T$ be the temperature of the cloud. 
The cloud is bathed in an ionising radiation field, 
   of photon number density $n_\gamma$, 
   and the radiation will photoionise 
   some of the material in the cloud.  
With the presence of co-existing ions and neutral material, 
 CX reactions will operate, 
 and the CX reactions will in turn compete with the photoionisation process.  
In a back-of-the-envelope estimation, the condition that CX dominates 
  over photoionisation is simply  
\begin{equation} 
  \varsigma\;\! {n_{\rm H}}^2 \langle v \rangle\;\!  \sigma_{\rm cx} 
  >  n_\gamma n_{\rm H} c \;\!  \sigma_{\rm pi} \ , 
\end{equation} 
  where $\sigma_{\rm cx}$ is the CX cross-section, $\sigma_{\rm pi}$ is the photoionisation cross-section, 
    $c$ is the speed of light, and $\langle v \rangle$ is the effective velocity of the collision between the hydrogen and metal atoms/ions. 
    
Assuming energy equipartition for the species, we can obtain the estimates of thermal velocities of the atoms: 
\begin{equation}  
 \frac{3}{2}\;\! k_{\rm B} T  =  \frac{1}{2}\;\! m_{\rm i} {v_{\rm i}}^2   \  \  \  (\ {\rm i} = 1,\ 2,\ 3,\ \cdots \  ) \ ,   
\end{equation}  
  where $k_{\rm B}$ is the Boltzmann constant. 
The velocity of an atomic species with an atomic mass $A_{\rm i}$ 
 is therefore   
\begin{equation} 
 v_{\rm i}  =  \sqrt{\frac{3\;\! k_{\rm B} T}{A_{\rm i} \; \! m_{\rm H}}} 
    =  
    1.57 \times 10^6  A_{\rm i}^{-1/2} \left(  \frac{T}{10^4\;\! {\rm K}} \right)^{1/2} {\rm cm~s}^{-1} \ .  
\end{equation}    
For a two-species collision, we may set the effective velocity  
\begin{equation}
 \langle v \rangle   
   \approx  \langle v_{\rm ij} \rangle \equiv
   \frac{m_{\rm i}v_{\rm i} + m_{\rm j} v_{\rm j} }{m_{\rm i} + m_{\rm j}} \    
\end{equation}
   (using the centre-of-momentum frame).

Without losing generality, take all the metals as oxygen, with atomic mass $A = 16$. 
Then, we have $v = 1.57 \times 10^6\;\! {\rm cm~s}^{-1}$ for hydrogen and $0.39 \times 10^6\;\! {\rm cm~s}^{-1}$ for oxygen 
  at $T = 10^4\;\! {\rm K}$.  
Hence, the effective velocity $\langle v \rangle \approx 0.46 \times 10^6\;\! (T/ 10^4\;\! {\rm K}  )^{1/2}\;\! {\rm cm~s}^{-1}$.  
The atomic fraction of hydrogen is about 910,000 parts per million and that of oxygen is about 477 parts per million in the solar system 
 \citep[see e.g.][]{Arnett1996book}, 
  i.e., $\varsigma_{\rm o, \odot} \approx 0.52 \times 10^{-3}$. 
The criterion for CX dominating over photoionisation is therefore 
\begin{align}  
  n_{\rm H} & >   \frac{n_\gamma c}{\varsigma_{\rm o} \langle v \rangle}    \left( \frac{\sigma_{\rm pi}}{\sigma_{\rm cx}}\right) \nonumber  \\ 
    & =  {\tilde n}_{\rm H} 
    \left( \frac{n_\gamma }{0.1 \;\!{\rm cm}^{-3}}\right) \left( \frac{\varsigma_{\rm o,\odot}}{\varsigma_{\rm o}}  \right) 
       \left(\frac{ 10^4\;\!{\rm K}}{T} \right)^{1/2} 
        \left( \frac{\sigma_{\rm pi}}{6.33 \times 10^{-18}\;\!{\rm cm}^2} \right) \nonumber \\ 
        &  \ \ \ \ \  \times \left( \frac{10^{-15}\;\!{\rm cm}^2}{\sigma_{\rm cx,o}} \right)      \ ,  
\end{align} 
   where ${\tilde n}_{\rm H} =  7.94 \times 10^4\;\! {\rm cm}^{-3}$. 
In terms of (penetrative) column density, we have  
\begin{align} 
 N_{\rm H} & >  
  {\tilde N}_{\rm H}\;\! \left( \frac{\Delta \ell}{1\;\!{\rm pc}} \right)  \left( \frac{n_\gamma }{0.1 \;\!{\rm cm}^{-3}}\right) 
  \left( \frac{\varsigma_{\rm o,\odot}}{\varsigma_{\rm o}}  \right) 
       \left(\frac{ 10^4\;\!{\rm K}}{T} \right)^{1/2}  \nonumber  \\ 
   & \ \ \ \ \ \times  
        \left( \frac{\sigma_{\rm pi}}{6.33 \times 10^{-18}\;\!{\rm cm}^2} \right) \left( \frac{10^{-15}\;\!{\rm cm}^2}{\sigma_{\rm cx,o}} \right)    \ , 
\end{align}   
  where $\Delta \ell$ is the depth for the neutral hydrogen column density, 
  and ${\tilde N}_{\rm H} \approx 2.45 \times 10^{23} \;\!{\rm cm}^{-2}$ is the reference column density.

We may take $\sigma_{\rm pi} =  \sigma^{\rm [k]}_{\rm pi}$, 
  the Kramers cross-section, that is, 
\begin{equation}  
  \sigma^{\rm [k]}_{\rm pi}  = 6.33 \times 10^{-18} Z^{-2}\ \left(\frac{E_\gamma}{13.6\;\!{\rm eV}}\right)^{-3}\;\! {\rm cm}^2 \ ,  
\end{equation}  
 where $E_\gamma$ is the energy of the photoionisation photons, and $Z$ is the effective charge of the nucleus. 
For a starburst galaxy similar to M82, the production rate of ionisation photons is $\sim 10^{52}\;\! {\rm s}^{-1}$, 
  corresponding to a radiative power of about $10^{41}{\rm erg~s}^{-1}$ 
  \citep[see Fig.1 in][]{Leitherer2005AIPC}.  
The ionisation photon number density in its galactic outflow is roughly of the order of $0.1\;\! {\rm cm}^{-3}$. 
In order for CX to operate efficiently, the cloud needs to have a neutral hydrogen density significantly higher than $10^4\;\! {\rm cm}^{-3}$, 
   unless the ionisation-photon number density is strongly suppressed, 
   or alternatively, the ``metal'' content is anomalously high (which is unlikely).  
 Thus, the cold dense clumps are much more favourable than the geometrically thinner, less dense and less opaque H$\alpha$ filaments   
    as the sites for the production of CX emission.

Note that we have used $T \approx 10^4$~K 
  for the velocity estimates for the atoms and ions 
  in the dense clumps.  
The temperature  
  would be higher 
  if the free electrons, 
  which are products of photoionisation,  
  transfer some of their energy 
  to the atoms and ions through collisions.

Finally, we remark on additional conditions 
  for the survival of the H$\alpha$ emitting filaments. 
Visual inspection of the images of 
  galactic outflows, such as those in M82, 
  indicates that filaments 
  could have lengths of order 100~pc. 
Theoretical modelling of galactic outflows 
 driven by starbursts, 
 e.g. \cite{Chevalier1985}, 
 has shown the terminal speed of a flow 
 can reach as high as $700~{\rm km~s}^{-1}$. 
Suppose we take an initial discrepancy of, 
  say $\sim 500~{\rm km~s}^{-1}$,  
  between the speed of a clump and the material around it.  
It would then take less than 1~Myr 
  to stretch a filament 
  to a length of 100~pc
  from the torn off material from the cold dense clump. 
Such a filament would become a H$\alpha$ emitter 
  if it is warmed up to $10^4$~K 
  on a similar or shorter timescale. 
A further condition is that 
  the filaments are not vaporised 
  or disrupted during their warming process. 
This would require that 
  the evaporation timescale and 
  the turbulence disruption time scale 
  of the filament be greater than 1~Myr.


\bsp	
\label{lastpage}
\end{document}